\documentclass[preprint,superscriptaddress,floatfix,longbibliography,citeautoscript]{revtex4-2}

\usepackage{subcaption}

\usepackage{bibunits}
\defaultbibliographystyle{apsrev4-2}
\defaultbibliography{comboBib}

\usepackage{amsmath,amssymb} 
\usepackage{bm} 
\usepackage{graphicx} 
\usepackage{comment} 
\usepackage[normalem]{ulem} 
\usepackage{setspace}

\usepackage{minted} 
\usepackage{textcomp} 

\usepackage{enumitem}
\usepackage{xspace}

\setlist{noitemsep,leftmargin=*,topsep=0pt,parsep=0pt}

\usepackage{xcolor} 
\usepackage{float}
\definecolor{lightgray}{gray}{0.6}
\definecolor{medgray}{gray}{0.4}
\definecolor{mRed}{RGB}{230, 0, 50}
\colorlet{newtextColor}{mRed}
\makeatletter
\renewcommand{\fnum@figure}{\textbf{Fig.~\thefigure}}
\makeatother

\usepackage[hidelinks]{hyperref}
\hypersetup{
colorlinks=true,
urlcolor= blue,
citecolor=blue,
linkcolor= magenta,
}

\newif\ifptitle
\newif\ifpnumber
\newcounter{para}

\newif\iftrackchanges

\usepackage{mdframed}
\newmdenv[
  linecolor={\iftrackchanges newtextColor\else white\fi},
  linewidth=2pt,
  topline=false,
  bottomline=false,
  rightline=false,
  skipabove=\topsep,
  skipbelow=\topsep,
  leftmargin=-12pt,
  innertopmargin=0pt,
  innerbottommargin=0pt
]{newtextblock}

\ptitletrue  
\pnumbertrue 

\newcommand{\LNO}{La$_{3}$Ni$_{2}$O$_7$\xspace}

\newcommand{\Tc}{$T_\textrm{c}$\xspace}

\newcommand{\Lafour}{La$_4$Ni$_3$O$_{10}$\xspace}
\newcommand{\Tdw}{$T_\textrm{DW}$\xspace}

\newcommand{\harvardphys}{Department of Physics, Harvard University, Cambridge, MA 02138, USA}

\newcommand{\harvardeng}{John A. Paulson School of Engineering and Applied Sciences, Harvard University, Cambridge, MA 02138, USA}

\newcommand{\miteecs}{Department of Electrical Engineering \& Computer Science, Massachusetts Institute of Technology, Cambridge, MA 02139, USA}

\newcommand{\aps}{Advanced Photon Source, Argonne National Laboratory,  Lemont, IL 60439, USA}

\newcommand{\asu}{Department of Physics, Arizona State University, Tempe, AZ 85287, USA}

\newcommand{\aep}{School of Applied \& Engineering Physics, Cornell University, Ithaca, NY 14853, USA}

\newcommand{\kavli}{Kavli Institute at Cornell for Nanoscale Science, Cornell University, Ithaca, NY 14853, USA}

\newcommand{\mpi}{Max Planck Institute for Chemical Physics of Solids, 01187 Dresden, Germany}

\newcommand{\chess}{
Cornell High Energy Synchrotron Source, Cornell University, Ithaca, NY 14853, USA}

\newcommand{\mytitle}{Persistent structural distortions and absent superconductivity in trilayer nickelate thin films}

\begin{document}

\begin{bibunit}
\title{\mytitle}

\author{Abigail Y. Jiang}
\thanks{These authors contributed equally to this work.}
\affiliation{\harvardeng}
\affiliation{\harvardphys}

\author{Maria Bambrick-Santoyo}
\thanks{These authors contributed equally to this work.}
\affiliation{\harvardphys}
\affiliation{\miteecs}

\author{Lopa Bhatt}
\affiliation{\aep}

\author{Kyeong-Yoon Baek}
\affiliation{\harvardphys}

\author{Yi-Feng Zhao}
\affiliation{\asu}

\author{Dan Ferenc Segedin}
\affiliation{\harvardphys}

\author{Ari B. Turkiewicz}
\affiliation{\harvardphys}

\author{Jenna Hatmin}
\affiliation{\harvardphys}

\author{Grace A. Pan}
\affiliation{\harvardphys}


\author{Suchismita Sarker}
\affiliation{\chess}

\author{Donald A. Walko}
\affiliation{\aps}

\author{Charles M. Brooks}
\affiliation{\harvardphys}

\author{David A. Muller}
\affiliation{\aep}
\affiliation{\kavli}

\author{Berit H. Goodge}
\affiliation{\mpi}

\author{Hua Zhou}
\affiliation{\aps}

\author{Antia S. Botana}
\affiliation{\asu}

\author{Julia A. Mundy}
\email{mundy@fas.harvard.edu}
\affiliation{\harvardphys}
\affiliation{\harvardeng}

\date{\today}

\begin{abstract}
A new family of high-temperature superconductors was recently discovered in the $n=2,3$ Ruddlesden--Popper nickelates, where superconductivity emerges concomitant with suppression of parent density waves and structural octahedral rotations under hydrostatic pressure.
Intriguingly, compressive strain mimics the structural effects of pressure in the $n=2$ phase, yielding ambient-pressure superconductivity.
However, analogous strain-stabilized superconductivity has not been realized in the $n=3$.
Here, we use atomically-precise synthesis, transport, picoscale electron microscopy, and synchrotron X-ray diffraction to probe $n=3$ \Lafour thin films.
Although compressive strain suppresses density wave order, we do not observe superconductivity even under the largest strain state.
Importantly, we identify a structural distortion unique to strained $n=3$ thin films that may inhibit superconductivity: persistent, layer-inequivalent octahedral rotations around the $c$-axis.
Our results highlight key differences between the $n=3$ and $n=2$ systems, suggesting that ambient-pressure superconductivity in the $n=3$ may require new methods beyond epitaxial strain engineering.
\end{abstract}

\maketitle 


\newpage
\section*{Introduction}
Following the discovery of high-temperature (\Tc) superconductivity in cuprates and iron pnictides, there has been a long-standing effort to understand their underlying phenomena and engineer other superconducting systems \cite{Keimer2015_cuprateReview, Fernandes2022_FeSCreview}.
Nickelates have emerged as the latest class of superconductors \cite{Puphal2026_nickelatesReview, Goodge2025_nickelatesPerspective}: first in the square-planar phases \cite{Li2019_infiniteLayer, Pan2022_Nd6Ni5O12, Pan2026_dome}, 
and most recently in the high-\Tc Ruddlesden--Popper (RP) compounds under hydrostatic pressure \cite{Sun2023_La327,Li2024_La4310} and epitaxial strain \cite{Ko2024_La327ambient, Zhou2025_LaPr327films}. Although these materials families have notable distinctions across crystal and electronic structures, they share an important phenomenological feature: competition between the superconducting state and correlated density wave order 
\cite{Croft2014_CuCDW, Chang2012_YBCO2, Fang2009_FeMultiband, Yi2014_FeCompetition}.

In the RP nickelates ($R_{n+1}$Ni$_n$O$_{3n+1}$, $R$ = rare earth), pressure-driven competition between density waves and superconductivity occurs concomitant with underlying changes in crystal structure.
At ambient pressure, both the bilayer ($n=2$) and trilayer ($n=3$) bulk crystals occupy their lowest symmetry space groups, orthorhombic $Amam$ and monoclinic $P2_1/a$ respectively, which host various NiO$_6$ octahedral rotations \cite{Sun2023_La327,Zhu2024_La4310}. These structures also host parent density waves \cite{Chen2024_SpinLa327,Zhang2020_intertwinedDW}, with several characterizations in $n=3$ of intertwined, incommensurate spin and charge density modulation
\cite{Zhang2020_intertwinedDW, Samarakoon2023_bootstrappedSDW,Jia2026_La4310latticeCharge}. 
In both the bilayer and trilayer nickelates under hydrostatic pressure, superconductivity emerges soon after the density waves are suppressed \cite{Li2025_La327,Zhang2024_La327,Khasanov2025_La327Splitting,Zhang2025_La4310,Li2024_La4310,Li2024_La4310structure,Huang2024_Pr4310,Chen2025_Pr4310,Zhang2025_Pr4310}, 
showing remarkable similarity to the cuprate and iron-based superconducting phase diagrams \cite{Keimer2015_cuprateReview, Fernandes2022_FeSCreview}. Concomitant with stabilization of superconductivity, the $n=2,3$ nickelate transform to higher-symmetry tetragonal $I4/mmm$ crystal structures, where NiO$_6$ octahedral rotations are completely suppressed 
\cite{Wang2024_La327, Li2025_La327, Zhu2024_La4310, Zhang2025_La4310, Li2024_La4310structure}.
The stabilization of ambient-pressure superconductivity in compressively strained $n=2$ thin films \cite{Ko2024_La327ambient} enabled microscopy studies \cite{Bhatt2025_La327films}, which revealed shared structural features across distinct realizations of RP nickelate superconductivity. In particular, multislice electron ptychography (MEP) of superconducting bilayer \LNO thin films identified complete suppression of NiO$_6$ octahedral rotations, consistent with the bulk pressure-driven bilayer phase diagram \cite{Bhatt2025_La327films}.

\begin{figure*}[b]
\includegraphics[clip=true,width=\columnwidth]{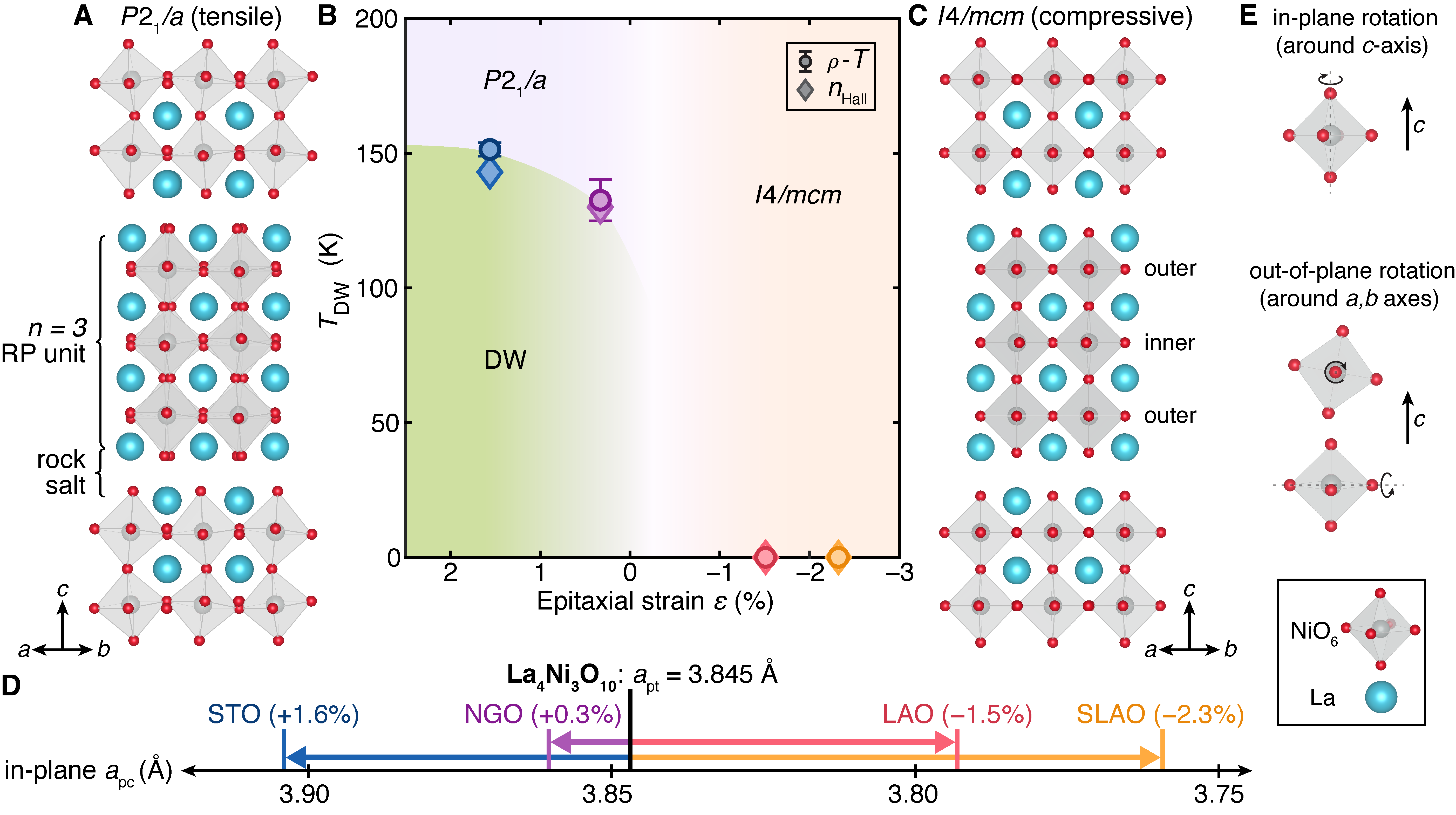}
    \caption{\textbf{Phase diagram of epitaxially strained \Lafour thin films.} 
    \textbf{a)} Crystal structure of \Lafour in the monoclinic $P2_1/a$ bulk space group \cite{Zhang2020_LaPr4310floatingzone} with pseudo-orthorhombic axes labels as observed in \Lafour under tensile strain.
    \textbf{b)} Electronic and structural evolution with respect to epitaxial strain ($\varepsilon$). Density wave (DW) transition temperatures \Tdw are determined from resistivity ($\rho$ -- $T$, circles) and carrier density ($n_{\mathrm{Hall}}$, diamonds). No superconductivity is observed. Structural space groups are determined by electron microscopy and synchrotron X-ray diffraction.
    \textbf{c)} Tetragonal space group $I4/mcm$ observed in \Lafour under compressive epitaxial strain. Inner vs. outer perovskite layers are labeled for reference.
    \textbf{d)} Number line comparing the epitaxial strain imparted on bulk pseudo-tetragonal (pt) \Lafour by pseudo-cubic (pc) substrates STO, NGO, LAO, and SLAO from comparison of in-plane lattice parameters. 
    The corresponding strain state of \Lafour on each substrate is denoted in parentheses. 
    \textbf{e)} Depiction of in-plane (around the $c$-axis) vs. out-of-plane (around the $a,b$ axes) NiO$_6$ octahedral rotations referenced throughout the text.
    }
    \label{fig:xtal_phase}
\end{figure*}

While several groups have since reported superconductivity in various bilayer thin films \cite{Ko2024_La327ambient,Zhou2025_LaPr327films,Hao2025_SrLa327films,Liu2025_La2Pr327transport,Tarn2026_bilayerStrain}, analogous ambient-pressure superconductivity has not yet been realized in the trilayer thin film counterpart. This lack of superconductivity is particularly puzzling given that $n=2$ and $n=3$ derive from the same structural family and exhibit strikingly similar phase diagrams with respect to pressure. 
Here, we synthesize a series of high-quality thin films of trilayer \Lafour over a wide range of epitaxial strains, from which we construct a strain-dependent electronic and structural phase diagram (Fig.~\ref{fig:xtal_phase}A-C) supported by first-principles calculations. We show that compressive epitaxial strain systematically suppresses the parent density wave in \Lafour thin films while raising crystal symmetry, seemingly analogous to hydrostatic pressure applied to \Lafour bulk crystals. However, despite complete density wave suppression, we do not observe signatures of superconductivity in these \Lafour thin films, even under larger compressive strains than those required to achieve superconductivity in \LNO  \cite{Ko2024_La327ambient} (Fig. \ref{fig:supp_straincomp}). Using a combination of electron microscopy and synchrotron X-ray diffraction, we identify remnant in-plane (i.e. around the $c$-axis) octahedral rotations unique to $n=3$ that persist under compressive strain
as a potential structural feature inhibiting emergence of superconductivity. In contrast to bilayer \LNO where compressive strain completely suppresses octahedral rotations, we observe an intermediate $I4/mcm$ structure in compressively strained \Lafour (Fig.~\ref{fig:xtal_phase}C) that falls just short of the highest-symmetry bulk superconducting $I4/mmm$ structure, revealing a strain-driven structural evolution of thin film \Lafour that is unique from bulk crystals under pressure.


\section*{Results}
We synthesize \Lafour thin films via reactive oxidative molecular-beam epitaxy (MBE)
controlled by precise shuttering as described in ~\cite{methods}. Using pseudo-cubic substrates, we stabilize four distinct epitaxial strain states ranging from tensile strain of $+1.6\%$ on SrTiO$_3$ (STO) and $+0.3\%$ on NdGaO$_3$ (NGO), to compressive strain of  $-1.5\%$ on LaAlO$_3$ (LAO) and $-2.3\%$ on SrLaAlO$_4$ (SLAO) (see Fig.~\ref{fig:xtal_phase}D). 
To preserve in-plane strain throughout each sample, films under moderate strain (STO, NGO, LAO) are 6 RP units ($\sim8-9$ nm) thick, while films on the largest magnitude strain (SLAO) are only 4 RP units ($<6$ nm) to remain below their critical thickness (Fig.~\ref{fig:supp_RSM_relaxation}).
We show reciprocal space maps (RSMs) demonstrating coherent strain across each sample in Fig.~\ref{fig:ext_RSM} and a strain-dependent evolution of the $c$-axis in Table \ref{tab:supp_structparams}.

Lab-based X-ray diffraction (XRD) patterns
show high quality thin films, with clear superlattice peaks arising from the $n=3$ RP layering structure, and Kiessig thickness fringes arising from clean interfaces (Fig.~\ref{fig:ext_labXRD}).
\textit{In-situ} reflection high energy electron diffraction (RHEED) and \textit{ex-situ} atomic force microscopy (AFM) also indicate highly crystalline films with smooth surfaces (Figs.~\ref{fig:supp_RHEED},~\ref{fig:supp_AFM}). Annular dark field scanning transmission electron microscopy (ADF-STEM) images show coherent $n=3$ ordering across the four strain states with limited intergrowths of other $n$ (Fig.~\ref{fig:supp_LargeFOV_ADF}). 

\section{Strain-driven suppression of parent density wave}
\begin{figure*}[!h]
    \includegraphics[clip=true,width=\columnwidth]{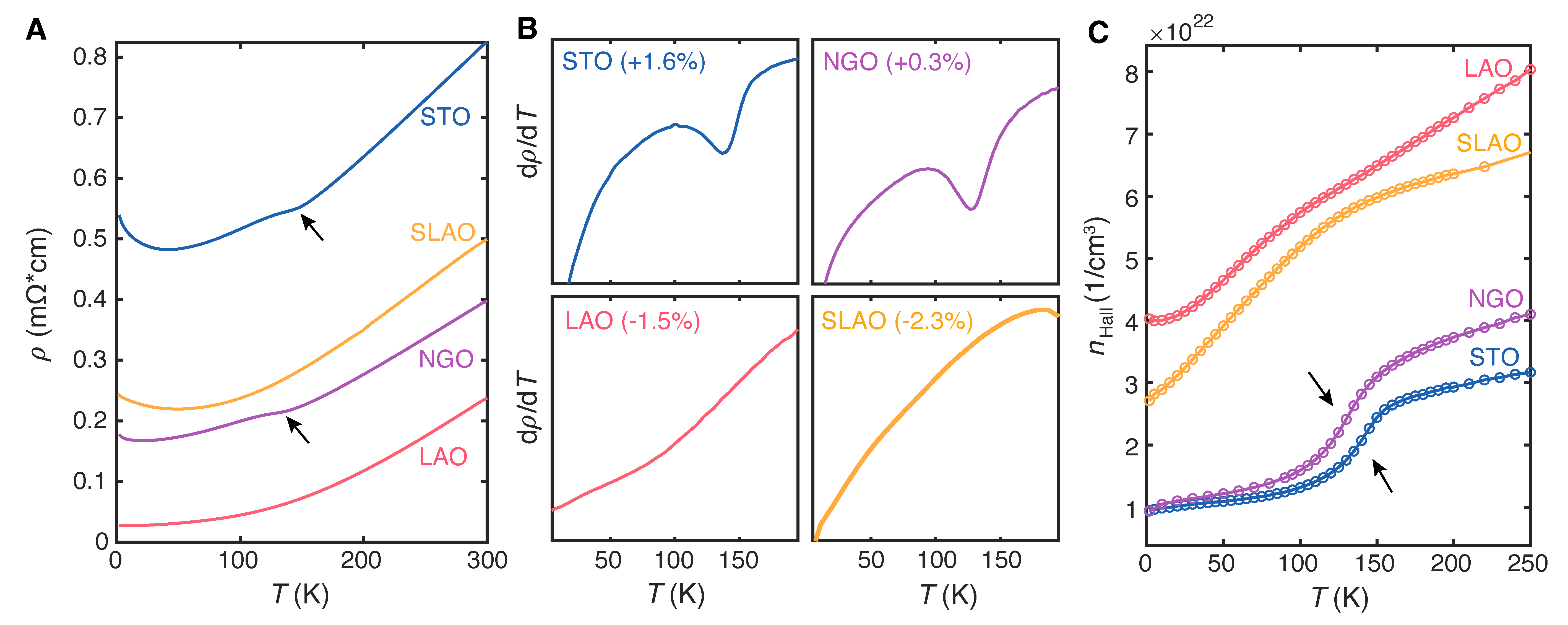}
    \caption{\textbf{Suppression of density wave under compressive strain in \Lafour thin films.} 
    \textbf{a)} Resistivity vs. temperature measurements across the strain series. The density wave metal-to-metal transition 
    at \Tdw are indicated by black arrows, present under tensile strain (NGO, STO) and absent under compressive strain (LAO, SLAO).
    \textbf{b)} First derivatives of the measurements in (a). A change in slope in the resistivity vs. temperature around \Tdw appears as a peak in the first derivative; this peak is present under tensile strain and absent under compressive strain in agreement with the resistivity curves.
    \textbf{c)} Temperature-dependent carrier density $n_{\text{Hall}}$, with abrupt changes in $n_{\text{Hall}}$ marked by black arrows occurring at similar temperatures to \Tdw from resistivity.}
     \label{fig:transport}
\end{figure*}

We use electrical transport to probe the density wave phase transition in our series of \Lafour thin films, from which we develop the strain-dependent phase diagram of electronic properties in Fig.~\ref{fig:xtal_phase}B. At ambient pressure, bulk \Lafour crystals display a characteristic metal-to-metal transition 
at $\sim$~135 K indicative of the density wave phase transition (\Tdw), which is suppressed under pressure prior to the emergence of superconductivity 
\cite{Zhang2024_La327,Khasanov2025_La327Splitting,Zhu2024_La4310,Li2024_La4310,Li2025_La4310structure}. 

Our resistivity measurements across the \Lafour strain series are shown in Fig.~\ref{fig:transport}A-B.
We observe a similar transition
in resistivity of \Lafour under the lowest strain state on NGO, with an average \Tdw~=~134~K (±~7~K) reproducibly over $>20$ samples (see \ref{sec:supp_transport}A and Figs. \ref{fig:supp_mainFig_derivs}-\ref{fig:ext_dome_allpoints} for more details). With increasing tensile strain on STO (+1.6\%), the transition temperature increases to an average \Tdw~=~152~K (±~3~K), suggesting that the density wave phase is more robust under increasing tensile strain. We also find that compressive epitaxial strain destabilizes the density wave, with no detectable transition
for samples on LAO ($-1.5$\%) and SLAO ($-2.3$\%). 
Sample statistics are detailed in Table \ref{tab:cdw_averages}. 

In a density wave system with a charge component, as identified in bulk \Lafour, the number of carriers at the Fermi energy is expected to decrease across the transition due to partial gapping of the Fermi surface. Hall measurements on the low-strain sample (\Lafour on NGO) are consistent with this picture: carrier density $n_{\mathrm{Hall}}$ exhibits an abrupt drop at 130~K, around the same temperature as \Tdw observed in resistivity (Fig.~\ref{fig:transport}C, \ref{sec:supp_transport}B).
On all other substrates, Hall measurement of the density wave are consistent with the observations from resistivity: tensile strain enhances \Tdw with the transition in $n_{\mathrm{Hall}}$ increased to 143~K on STO, while compressive strain suppresses the density wave with no 
abrupt transition observed in $n_{\mathrm{Hall}}$ on LAO or SLAO (Fig. \ref{fig:transport}C, Figs.~\ref{fig:supp_rxy}, \ref{fig:supp_Rh}).
First-principles calculations discussed in \ref{sec:DFT_mag} and Fig.~\ref{fig:en_mag} also corroborate our experimental destabilization of the density wave upon compressive strain. 


The shared effect of density wave suppression across pressurized bulk crystals and strained thin films suggests that compressive epitaxial strain may be a potential analogue to hydrostatic pressure in the trilayer RP nickelate. 
However, we do not observe any superconducting transition in \Lafour compressively strained on SLAO, despite spanning a wide range of oxygen stoichiometry over many samples (\ref{sec:supp_ozone} and Figs.~\ref{fig:supp_ozone_temp}-\ref{fig:supp_ozone_power}). We also do not observe superconductivity under even larger compressive strain states (Fig.~\ref{fig:supp_YAO}). This lack of superconductivity is particularly striking given that \Lafour on SLAO is already under more compressive strain than the superconducting bilayer counterpart 
\cite{Ko2024_La327ambient} (Fig.~\ref{fig:supp_straincomp}). Given the density wave suppression in compressively strained \Lafour thin films but striking lack of superconductivity, we turn to structural characterization to identify possible crystallographic features that underlie our observed electronic changes.

\section{Suppression of out-of-plane octahedral rotations under compressive strain}

\begin{figure*}[t]
    \includegraphics[clip=true, width=\columnwidth]{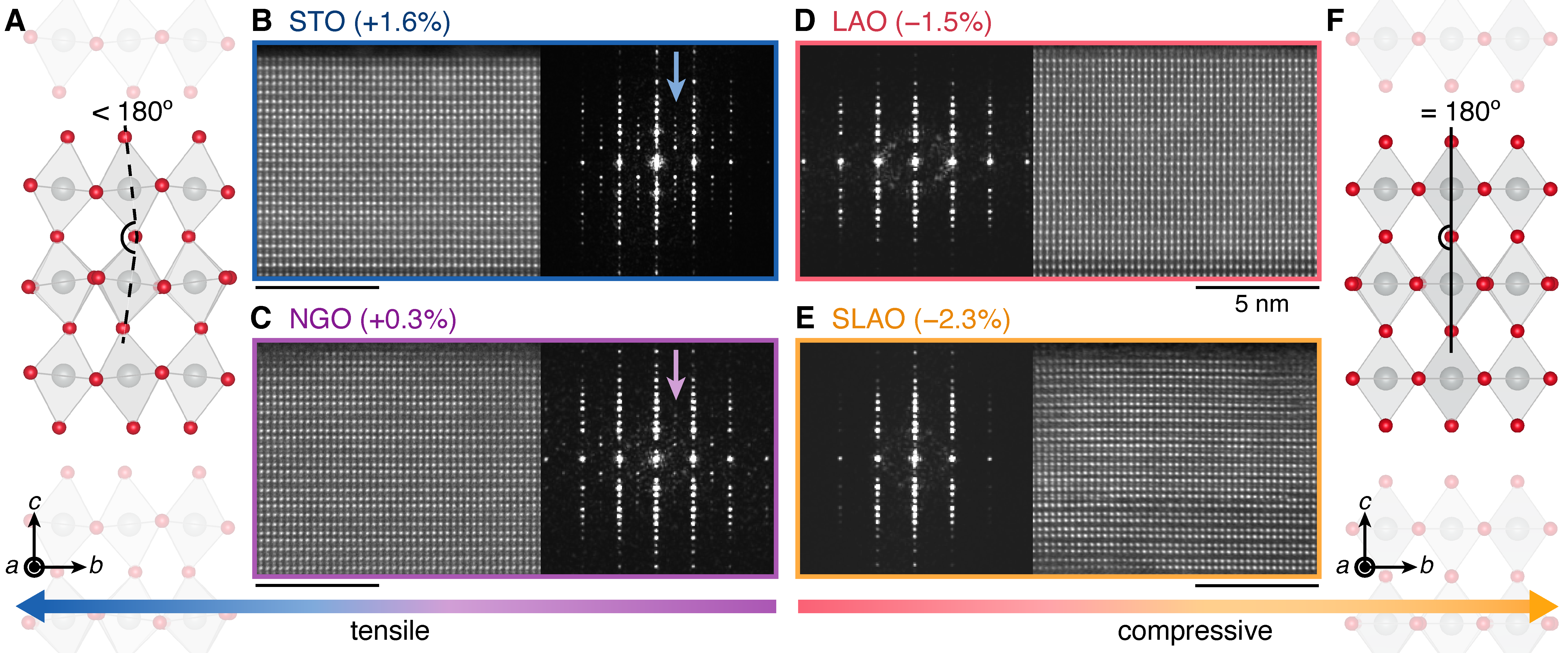}
    \caption{\textbf{Suppression of out-of-plane octahedral rotations and symmetry raising in \Lafour by compressive strain.} 
    \textbf{a)} NiO$_6$ octahedra of \Lafour in the low-symmetry $P2_1/a$ space group \cite{Zhang2020_LaPr4310floatingzone} projected along the orthorhombic \textit{a}-axis, [100]$_o$, representing the observed structure in tensile-strained films. The dashed lines indicate the distorted vertical Ni-O-Ni bond angle from out-of-plane octahedral rotations.
    \textbf{b-e)} ADF-STEM images projected along the orthorhombic $a$-axis [100]$_o$ of \Lafour thin films across the strain series, alongside Fourier transforms of each film region. Clear half-order peaks in the Fourier transforms (indicated by arrows) are observed in tensile-strained \Lafour. These peaks do not appear in compressively strained \Lafour, indicating a higher-symmetry structure without out-of-plane octahedral rotations stabilized by epitaxial compression.
    \textbf{f)} NiO$_6$ octahedra in the tetragonal structure adopted by compressively strained \Lafour,  where the solid line indicates vertical Ni-O-Ni bond angle straightening.}
     \label{fig:STEM}
\end{figure*}

We use a suite of structural characterization techniques to investigate changes in \Lafour structure and bonding, with particular focus on NiO$_6$ octahedral rotations, and correlate such changes to the films' electronic properties across strain states.
First, we use ADF-STEM to directly probe atomic positions and search for out-of-plane octahedral rotations (around $a,b$ axes, see Fig.~\ref{fig:xtal_phase}D).
These out-of-plane octahedral rotations and their corresponding vertical Ni-O-Ni bond angles have been widely discussed as important tuning knobs for nickelate RP electronic structure in both theory and experiment, where suppression of these rotations has been related to the emergence of superconductivity \cite{Sun2023_La327, Li2024_La4310structure, ZhaoBotana2025_RPstrain, Wang2024_La327, Li2025_La327, Bhatt2025_La327films}. 

We use Fourier transforms of ADF-STEM images projected along the orthorhombic $a$-axis, [100]$_o$, to probe such out-of-plane octahedral rotations from the film without contribution from the substrate or neighboring twinned film regions (Fig.~\ref{fig:supp_NGO_twin}). In an orthorhombic RP structure with 
out-of-plane rotations around only $a$, as seen in ambient pressure $n=3$ and $n=2$, we expect to observe half-order peaks in the Fourier transform projected along [100]$_o$ due to unit cell doubling along $b$.
\Lafour films under tensile strain (STO, NGO) exhibit clear half-order peaks (Fig.~\ref{fig:STEM}B,C) indicating presence of these out-of-plane rotations and vertical Ni-O-Ni bond angles smaller than 180\textdegree\ as visualized in Fig.~\ref{fig:STEM}A.
In contrast, for \Lafour under compressive strain (LAO, SLAO), the half-order peaks disappear (Fig.~\ref{fig:STEM}D,E), indicating suppression of out-of-plane rotations and straightening of vertical Ni-O-Ni bond angles to 180\textdegree\ as visualized in Fig.~\ref{fig:STEM}F. These measurements are repeatable across domains of \Lafour on all substrates, with large field-of-view images presented in Fig.~\ref{fig:supp_LargeFOV_ADF}. 
Out-of-plane octahedral rotation suppression upon compressive strain is also observed in our first-principles calculations, as discussed in \ref{sec:dft_struct}B.

Our observations of out-of-plane octahedral rotation suppression and vertical bond angle straightening on compressively strained \Lafour indicate transformation from orthorhombic to tetragonal symmetry, where the in-plane $a$ and $b$ axes become equivalent. 
This symmetry raising under compressive strain is also concomitant with density wave suppression observed in transport. Since the bulk \Lafour density wave propagates along orthorhombic $b^*$ \cite{Zhang2020_intertwinedDW}, its disappearance in this tetragonal phase is consistent with the loss of distinct $a$ and $b$ axes and thus the absence of a preferred $b^*$ ordering vector.
This out-of-plane bond angle straightening in $n=3$ is also consistent with $n=2$ thin films \cite{Bhatt2025_La327films}, suggesting that compressive strain is a shared method to 
quench out-of-plane octahedral rotations and thus 
raise crystal symmetry across epitaxial RP nickelates.

\section{Persistence of in-plane octahedral rotations under compressive strain}
\begin{figure*}[t]
    \includegraphics[clip=true,width=\columnwidth]{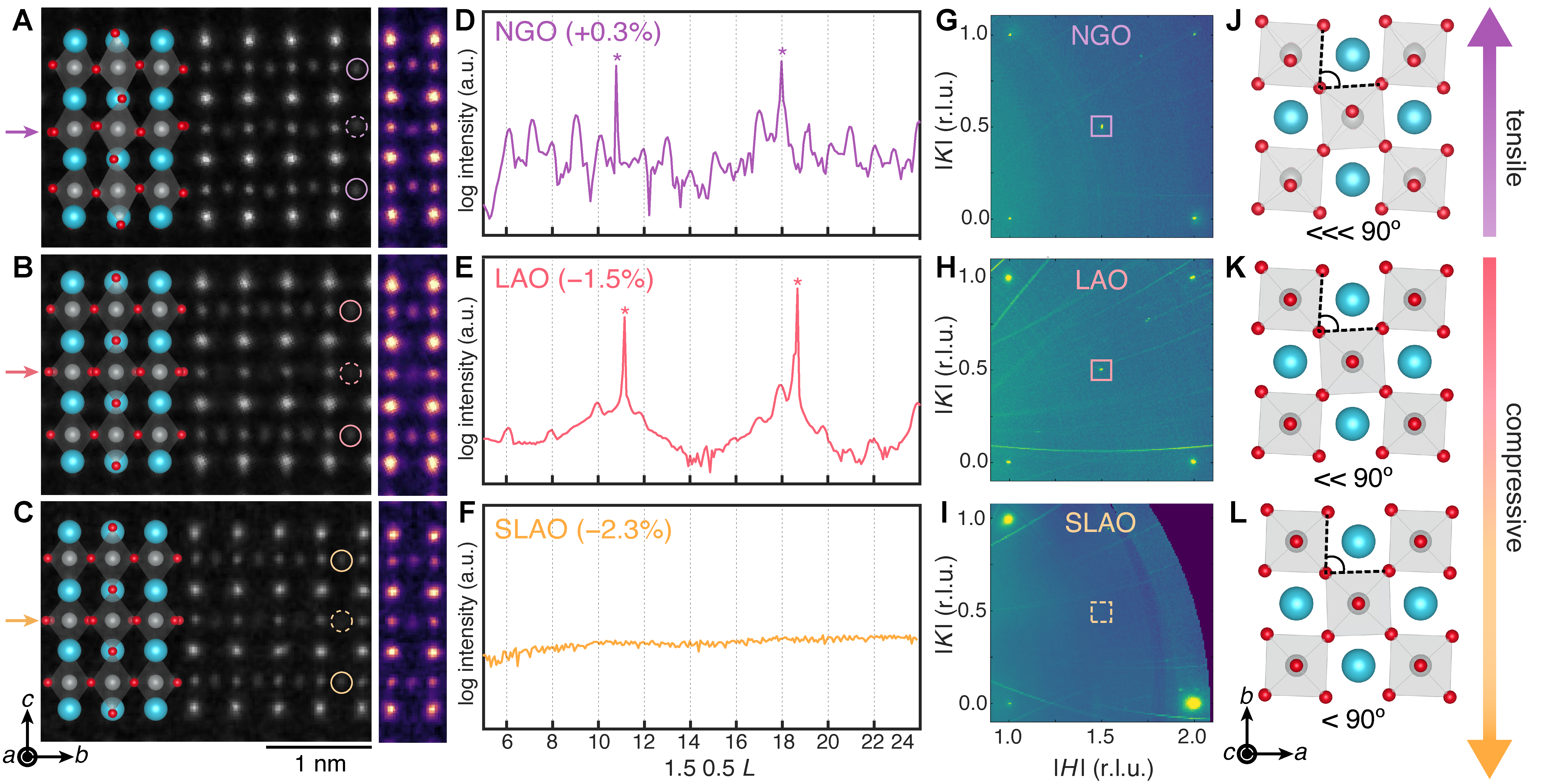}
    \caption{\textbf{Monotonic but incomplete suppression of in-plane rotations in \Lafour by compressive strain.} 
    \textbf{a)-c)} Ptychographic reconstructions of one $n=3$ RP unit projected along the orthorhombic \textit{a}-axis [100]$_o$ on NGO, LAO, and SLAO, overlaid with the corresponding atomic structures. The arrow indicates the inner layer of the RP unit which hosts in-plane rotations.
       Example sets of planar oxygen columns are labeled with circles: solid circles denote coherent planar O in the outer layers, in contrast to dotted circles that denote incoherent planar O columns in the inner layer.
    A saturated, magma-colored cropped field-of-view is presented on the right for better view of oxygen column contrast.
  \textbf{d)-f)} Synchrotron X-ray crystal truncation rods (CTRs) of strained \Lafour along ($H$=1.5, $K$=0.5, $L$) with respect to $L$, showing decrease in peak intensityfrom in-plane rotations upon compressive strain. Vertical dashed lines are at even $L$ as a guide to the eye; asterisks denote substrate peaks. The arrow marks the $L$ = 17 peak, corresponding to the squares in (g-i). 
  \textbf{g)-i)} $H$-$K$ plane cuts around (1.5 0.5 17) from synchrotron high dynamic-range reciprocal space mapping (HDRM).
    \textbf{j)-l)} Schematic depiction of the gradual but incomplete suppression of \Lafour in-plane octahedral rotations with increasing compressive strain.
    All data in this figure are indexed to the pseudo-tetragonal \Lafour unit cell.}
    \label{fig:COBRA}
\end{figure*}

In addition to out-of-plane rotations, the $n=3$ RP nickelate contains in-plane octahedral rotations (around the $c$-axis, see Fig.~\ref{fig:xtal_phase}D) that induce a monoclinic $P2_1/a$ structure at ambient pressure \Lafour \cite{Zhang2020_LaPr4310floatingzone}. 
Of particular importance is layer inequivalence within every trilayer RP unit, distinguished by differences between inner vs. outer layers as referenced in Fig. \ref{fig:xtal_phase}E. 
In the $P2_1/a$ structure, inner perovskite layers host in-plane rotations, while outer layers do not have in-plane rotations \cite{Zhang2020_intertwinedDW, Zhang2020_LaPr4310floatingzone}. These $n=3$ structural characteristics are distinct from the bulk $n=2$ system, which does not host any in-plane rotations \cite{Wang2025_La3Ni2O7_structure} and also does not have symmetry-related layer inequivalence.
Although less emphasized in literature compared to the out-of-plane Ni-O-Ni bond angle, in-plane bond angles have been theorized to play a key role in stabilizing superconductivity \cite{Samanta2025_bondAngles} but remain experimentally unreported in RP nickelate thin films.

We use cross-sectional multislice electron ptychography (MEP) 
to directly visualize light atoms (i.e. oxygen) at the picoscale in strained \Lafour thin films.
Along the same $a$-axis [100]$_o$ orthorhombic projection as ADF-STEM, ptychographic reconstructions reveal such an inequivalence between the inner vs. outer perovskite layers driven by in-plane octahedral rotations.
In MEP of \Lafour on NGO shown in Fig.~\ref{fig:COBRA}A, we observe that planar oxygen columns in the outer layers of the RP unit appear sharp and round (solid circle), while planar oxygen columns in the inner layer of the RP unit exhibit lower intensity and are more blurred (dotted circle). 
This preferential blurring arises from in-plane rotations that cause incoherent stacking of planar oxygen atoms along the beam direction in the inner layer only, as also denoted by the arrow in Fig \ref{fig:COBRA}A. 
Surprisingly, this inner-layer oxygen incoherence is present on all \Lafour samples, indicating the presence of in-plane rotations regardless of strain state (Fig.~\ref{fig:COBRA}B,C). 
These persistent layer-inequivalent rotations prevent the compressively strained \Lafour from reaching the bulk superconducting structure where all rotations are completely suppressed, even under the large magnitude of compressive strain imposed on \Lafour by SLAO ($-2.3$\%).  
Large field-of-view MEP on all substrates further corroborate these findings (Fig. \ref{fig:supp_LargeFOV_ptycho}).

Since MEP intensities cannot be quantitatively compared across different samples,
we turn to non-invasive synchrotron X-ray scattering as a complementary method to compare relative magnitudes of in-plane octahedral rotations at various strain states.
We use half-order crystal truncation rod (CTR) scattering, described in more detail in \ref{sec:supp_chess}A,
which is well-studied in traditional $AB$O$_3$ perovskites to map octahedral rotations \cite{May2025_bookChapter}. The (1.5 0.5 $L$) CTR specifically probes in-plane rotations \cite{Park2025_strainOOS, Adams2023_dynamicTilts}. 
In the (1.5 0.5 $L$) CTR of \Lafour on NGO, clear and intense peaks at all $L$ indicate presence of relatively large in-plane rotations (Fig.~\ref{fig:COBRA}D).
These peak intensities are suppressed with increasing compressive strain, with weakened intensity under moderate compressive strain (LAO, Fig.~\ref{fig:COBRA}E) and no detected intensity under large compressive strain (SLAO, Fig.~\ref{fig:COBRA}F). 
These CTR results are also cross-verified by synchrotron high dynamic-range reciprocal space mapping (HDRM) shown in Fig.~\ref{fig:COBRA}G-I, where the (1.5 0.5 $L$) peak intensity at a particular $L$=17 is suppressed with compressive strain. Further HDRM analysis details, including cuts at other $L$, are included in \ref{sec:supp_chess}B-D.
These weakening diffraction intensities indicate that in-plane octahedral rotations monotonically decrease in magnitude with increasing compressive strain.

Taken together, MEP and synchrotron X-ray measurements suggest that in-plane octahedral rotations gradually become less distorted (i.e. bond angles closer to 90\textdegree) under compressive strain, but are never fully suppressed even on SLAO (Fig.~\ref{fig:COBRA}J-L). 
Our ptychographic reconstructions also highlight inequivalence within the $n=3$ RP unit, where in-plane rotations remain confined to the inner layer across all strain states.
These persistent in-plane rotations -- also found in first-principles structural calculations discussed in \ref{sec:dft_struct} -- mark a surprising departure from the picture in bilayer \LNO, 
where octahedral rotations are completely suppressed via compressive strain 
\cite{Bhatt2025_La327films}. 
Instead, these in-plane rotations in $n=3$ prevent strained \Lafour from reaching the highest symmetry structure found to superconduct in bulk, emphasizing the subtle but important effects of $n$ on crystal symmetry and electronic structure in RP nickelates.

\section{Structural evolution of \texorpdfstring{\Lafour}{La4Ni3O10} with strain vs. pressure}

\begin{figure*}[t]
    \includegraphics[clip=true,width=0.5\columnwidth]{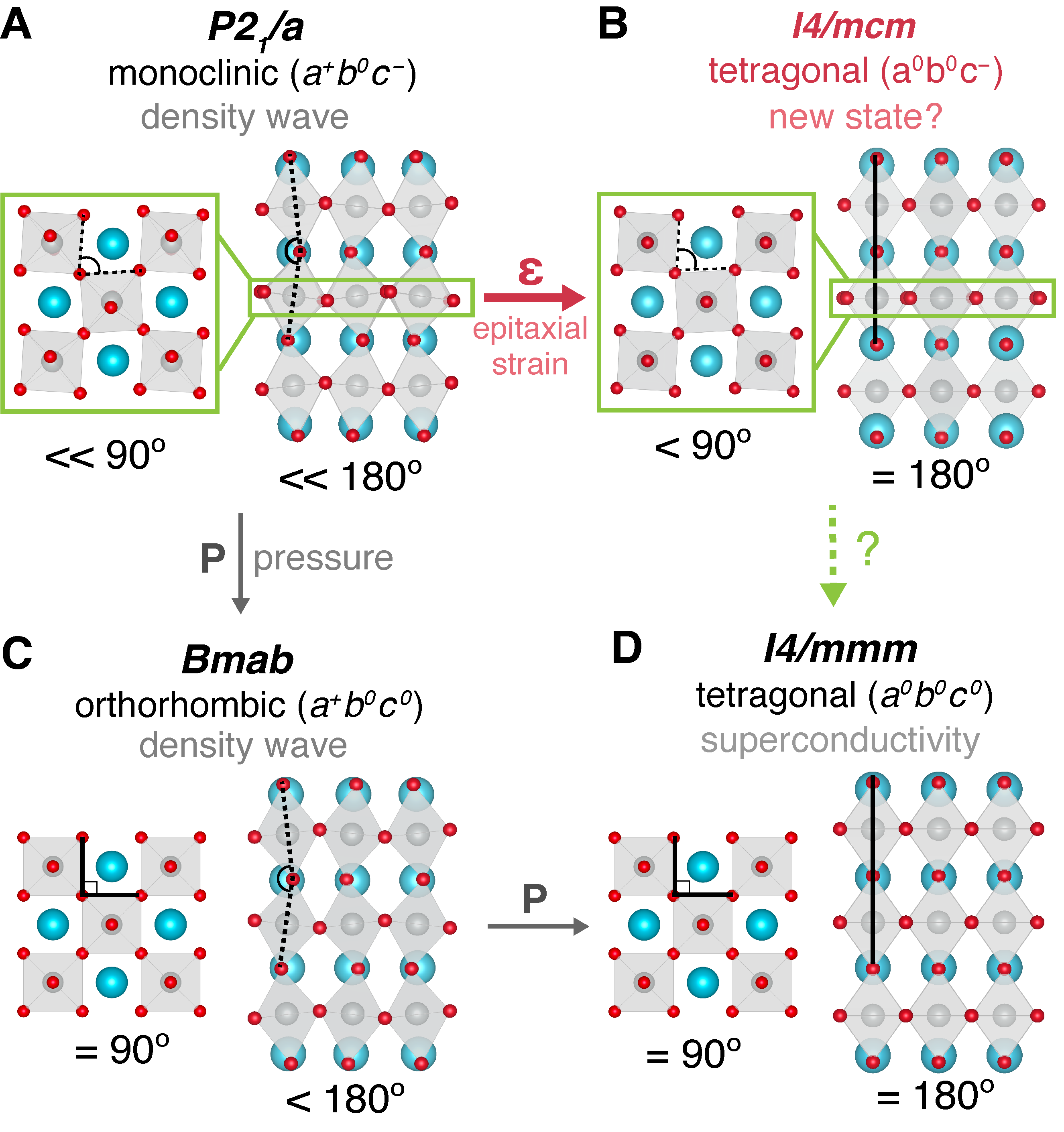}
    \caption{\textbf{Evolution of \Lafour crystal structure under bulk pressure vs. epitaxial strain.}} 
    \textbf{a)} Monoclinic $P2_1/a$ symmetry found in ambient pressure bulk crystals \cite{Zhang2020_LaPr4310floatingzone} and tensile-strained \Lafour thin films, which hosts both in-plane and out-of-plane rotations. 
    \textbf{b)} Tetragonal $I4/mcm$ structure stabilized under compressive epitaxial strain, with suppressed out-of-plane rotations but remnant in-plane rotations.
    \textbf{c)} Orthorhombic $Bmab$ structure stabilized under intermediate pressures in bulk 
    \cite{Nagell2017_LaCo4310}, with out-of-plane but not in-plane rotations. 
    \textbf{d)} $I4/mmm$ tetragonal symmetry \cite{Nagell2017_LaCo4310}found to be superconducting in bulk, with no octahedral rotations. 
    Full crystal structures of all four \Lafour space groups are presented in \ref{sec:supp_structures}.
     \label{fig:structure}
\end{figure*}

To make structural comparisons between bulk and thin film \Lafour, we first inspect the octahedral rotations that underlie crystal symmetry and electronic evolutions in bulk \Lafour under hydrostatic pressure.
At ambient pressure, bulk \Lafour exhibits density wave order and adopts the $P2_1/a$ monoclinic space group. This $P2_1/a$ structure hosts out-of-plane rotations, as well as in-plane rotations that are present in the inner layer but absent in outer layers \cite{Zhang2020_LaPr4310floatingzone} (Fig.~\ref{fig:structure}A). These rotations correspond to $a^+b^0c^-$ in Glazer notation, commonly used to index octahedral rotations in perovskites \cite{Woodward1997_tiltingI,Howard1998_GroupTheory} (see Fig. \ref{fig:supp_p21a} for more detail). At intermediate pressures where the density wave is pushed to lower temperature, the bulk \Lafour space group evolves from monoclinic $P2_1/a$ to orthorhombic $Bmab$ (Fig.~\ref{fig:structure}C) as identified by single-crystal diffraction studies \cite{Li2025_La4310structure}. This $Bmab$ structure has no in-plane rotations but retains out-of-plane rotations, corresponding to $a^+b^0c^0$ (Fig. \ref{fig:supp_bmab}). Under pressures of approximately 15-20 GPa \cite{Li2025_La4310structure, Li2024_La4310, Zhu2024_La4310, Zhang2025_La4310, Li2025_La4310XAS}, bulk \Lafour undergoes a structural transition to the highest-symmetry tetragonal $I4/mmm$ space group where all rotations are quenched corresponding to $a^0b^0c^0$ (Fig.~\ref{fig:structure}D, Fig. \ref{fig:supp_i4mmm}). This tetragonal $I4/mmm$ structure appears prerequisite to density wave suppression and emergence of superconductivity \cite{Li2025_La4310structure, Li2024_La4310, Zhu2024_La4310, Zhang2025_La4310}.

Next, we assign octahedral rotation patterns and their corresponding space groups to our \Lafour thin films at different strain states. Our structural evidence suggests that \Lafour under tensile strain on both NGO and STO adopt the same rotations observed in the $P2_1/a$ bulk structure shown in Fig.~\ref{fig:structure}A. These tensile-strained films host both in-plane and out-of-plane rotations corresponding to  $a^+b^0c^-$, with in-plane rotations remain confined to the inner perovskite layer, in agreement with the bulk $P2_1/a$ structure.
For \Lafour under larger tensile strain on STO, the higher \Tdw observed via transport and Hall is consistent with more distorted octahedra compared to the smaller strain state of \Lafour on NGO \cite{Norman2025_LandauDW, Rout2020_R4310structure}, but remains in this same $P2_1/a$-like structure.

Our characterization of \Lafour on LAO and SLAO suggests a distinct structural transition for strained thin films compared to pressurized bulk crystals. In marked contrast to pressure, compressive strain in the trilayer system instead first suppresses out-of-plane octahedral rotations while retaining in-plane rotations. These compressively strained \Lafour thin films host $a^0b^0c^-$ rotations (also denoted as $a^0a^0c^-$ by tetragonality), retaining inner layer in-plane rotations as visualized in Fig.~\ref{fig:structure}B and Fig. \ref{fig:supp_i4mcm}, with decreasing bond distortion magnitudes under increasing compressive strain. 
Based on these observed rotations and without evidence of further complex distortions (e.g. cation displacements), we assign a highest possible space group of $I4/mcm$ to compressively strained \Lafour from Glazer notation \cite{Howard1998_GroupTheory}. This $I4/mcm$ symmetry is notably not observed in bulk, and indicates a new state and structural evolution pathway for \Lafour uniquely driven by strain.
Although the density wave appears fully suppressed via electronic transport, these compressively strained \Lafour thin films do not yet reach the highest-symmetry, bulk superconducting $I4/mmm$ structure with all rotations suppressed. Given that we reach practical limits of traditional epitaxy due to synthetic challenges (e.g. rapid strain relaxation, Fig.~\ref{fig:supp_YAO}) on even larger compressive strain states as discussed, possible pathways to bridge the $I4/mcm$ and $I4/mmm$ structures remain an open question as indicated in Fig. \ref{fig:structure}.

\section{Conclusion and Outlook}

Here, we identify critical distinctions between the effects of compressive epitaxial strain and hydrostatic pressure on competing electronic phases and their underlying crystal structures in trilayer \Lafour. Compared to hydrostatic pressure applied to bulk crystals, compressive strain applied to thin films mimics two main effects of pressure on \Lafour beyond shrinking in-plane lattice constants: suppression of the parent density wave and straightening of the vertical Ni-O-Ni bond.
However, we find that compressive epitaxial strain is not a complete analogue for
hydrostatic pressure in the trilayer RP nickelate. Instead, it drives a unique structural evolution where out-of-plane rotations are first quenched but in-plane rotations persist -- even under larger strains than needed to realize superconductivity in bilayer thin films. 
This series of rotations defines a previously unobserved structure in RP nickelates, the $I4/mcm$ space group, demonstrating the capability of epitaxial strain to induce unique structural variants in correlated oxides. 
Our observations may also contextualize the lack of superconductivity in compressively strained ``1313'' superlattice RP nickelate thin films in contrast to other $n=2$ - containing structural variants such as the ``1212",
\cite{Nie2026_superstructures} despite observations of superconductivity in bulk pressurized ``1313'' \cite{Chen2024_1313, Puphal2024_1313}.

The persistence of in-plane octahedral rotations in compressively strained \Lafour marks a 
sharp contrast to bilayer \LNO, where both compressive strain and hydrostatic pressure appear to fully suppress all octahedral rotations and simultaneously stabilize superconductivity (Fig.~\ref{fig:ext_La327_structure}).
The inherent symmetry difference between inner and outer perovskite layers in the $n=3$ RP nickelate 
affords an additional structural degree of freedom and thus allows persistent rotations confined to the inner layer, unlike $n=2$ where all layers are structurally equivalent by symmetry.
Theoretical calculations also emphasize this layer differentiation through magnetic inequivalence between the inner and outer layers in \Lafour \cite{LaBollita2024_La4310, AuYeung2025_universalNickelates}, highlighting the importance of disentangling layer-resolved electronic contributions in the $n=3$ and higher-order RP structures. Additional experimental probes such as layer-resolved ARPES could further contextualize the role of inner vs. outer layers in density wave suppression and possible stabilization of superconductivity. Such studies on RP nickelates would be particularly interesting in comparison to the cuprates, where the highest \Tc s derive from layer inequivalence in the trilayer systems \cite{Luo2023_trilayerCuprates, Kornilovitch2026_Tcmax}. 


Our systematic study also opens new questions about how to engineer ambient-pressure superconductivity in trilayer nickelate compounds. Although reaching larger compressive strain states is a tempting method to potentially stabilize superconductivity in $n=3$ thin films, our work demonstrates the limitations
of compressive strain as a method to stabilize high-symmetry $I4/mmm$ \Lafour. Instead, new synthesis avenues such as complex layering strategies that leverage layer inequivalences within the $n=3$ structure, or otherwise promote bulk-like trilayer electronic structure, may be key to ambient-pressure superconductivity in trilayer RP nickelates.


\section*{Methods}

\subsection*{Thin film synthesis} 
All samples were synthesized with reactive ozone molecular beam epitaxy (MBE). The La and Ni flux were first determined by synthesis of binary oxides \cite{Sun2022_fluxCal}, and flux-matched by optimization of stoichiometric LaNiO$_3$ films. \textit{In-situ} RHEED monitoring and \textit{ex-situ} X-ray reflectivity measurements were used to extract monolayer La and Ni shutter times from LaNiO$_3$, similar to the methods discussed in \cite{Pan2022_NdRP, Segedin2023_strainLimits}. Using these shutter times and the dynamic layer-by-layer shuttering scheme proposed by \cite{Nie2014_nonstoich,Lee2014_dynamicLayer} and previously demonstrated in \cite{Pan2026_dome}, \Lafour films were synthesized at 500-600\textdegree C pyrometer temperature and in 1.3 × 10$^{-6}$ torr distilled ozone on four pseudo-cubic substrates: SrLaAlO$_4$ (001), LaAlO$_3$ (001), NdGaO$_3$ (110), and SrTiO$_3$ (100).

Substrates are purchased from OST Photonics, CrysTec, Shinkosha, and MTI Corporation. Before MBE synthesis, STO, NGO, and LAO substrates were cleaned and annealed in air to achieve atomic step terraces. SLAO substrates were not annealed in air prior to synthesis due to crystal degradation; see Figs. \ref{fig:supp_SLAO_XRD},\ref{fig:supp_SLAO_AFM} for additional details. For all substrates, 200 nm Pt was deposited via e-beam on the rough back-side before use in the MBE to promote uniform thermalization during thin film synthesis. Films on SLAO were annealed post-MBE synthesis in ozone to promote filling of oxygen vacancies (\ref{sec:supp_ozone} and Figs.~\ref{fig:supp_ozone_temp}-\ref{fig:supp_ozone_power}).

\subsection*{Electrical transport measurements}
Transport measurements were conducted in a Quantum Design Dynacool Physical Property Measurement System (PPMS) equipped with a 9T magnet. All measurements were collected at 6-12 Hz AC current. For resistivity vs. temperature and Hall measurements, 20 nm Pt or Pd contacts were deposited via e-beam on each sample and ultrasonically wirebonded using Si-doped Al wire. Additional resistance vs. temperature measurements were conducted by direct wirebonding to the sample surfaces. Hall measurements were all taken upon warming, and coefficients $R_H$ were fit from anti-symmetrized field sweeps (\ref{sec:supp_transport}B and Figs.~\ref{fig:supp_rxy}, ~\ref{fig:supp_Rh}).

\subsection*{Electron microscopy and multislice electron ptychography}
To prepare cross-sectional lamellas, we used a standard focused ion beam (FIB) lift-out procedure using Tescan Amber X Plasma FIB or a Thermo Fisher Helios G4 UX FIB.  We performed STEM imaging in ADF mode using a Cs-corrected Thermo Fisher Scientific (TFS) Spectra 300 X-CFEG or aberration-corrected TFS Titan Themis 300 operating at 300 kV with probe convergence semi-angles of 30 mrad and 21.4 mrad, respectively.

Ptychographic 4D-STEM datasets were collected on the same tools equipped with an EMPAD-G2 detector \cite{EMPAD2}.
The maximum likelihood algorithm implemented in the fold-slice package was used to perform iterative phase retrieval \cite{ZhenMEPScience,Thibault2012Maximumlikelyhood,Wakonig2020Ptychoshelves}, using position correction and multiple probe modes to account for partial coherence of the STEM probe \cite{Thibault2013Mixprobe,zhen2020MixedState}. 
A Bayesian optimization algorithm with data error as the objective function was used to optimize parameters such as defocus, convergence angle, and rotation angle \cite{chenyu2022BO}.

\subsection*{First-principles calculations}
We used density functional theory (DFT) calculations as implemented in QUANTUM ESPRESSO \cite{Giannozzi_2009} to investigate the structural and magnetic properties of \Lafour. The chosen exchange-correlation functional was the Perdew-Burke-Ernzerhof (PBE) parametrization of the generalized gradient approximation (GGA) \cite{perdew1996generalized}. For the strained structures, we fixed the in-plane lattice constants and fully relaxed the out-of-plane lattice constant as well as internal coordinates within a nonmagnetic state. The energy cut-off for the plane wave basis was set to 80 Ry. The Monkhorst-Pack k-point grid was used with a sampling of 8$\times$8$\times$3 for the $P2_1/a$ phase, 7$\times$7$\times$2 for $Bmab$, 11$\times$11$\times$2 for $I4/mmm$, and 6 $\times$6$\times$2 for $I4/mcm$, respectively (see \ref{sec:dft_struct} for further details). To estimate the magnetic ground state of \Lafour at ambient pressure and under strain, we employed the GGA+$U$ method \cite{dudarev1998electron}, with a $U$ parameter ranging from 2 eV to 6 eV. The magnetic configurations used were ferromagnetic (FM), A-type antiferromagnetic (A-AFM), C-type antiferromagnetic (C-AFM), G-type antiferromagnetic (G-AFM), and a so-called M/0/M state to mimic the experimental density wave (see \ref{sec:DFT_mag} for further details). 

\subsection*{Lab-based X-ray diffraction}
In-house X-ray diffraction (XRD) measurements were conducted with a Malvern Panalytical Empyrean diffractometer with Cu K$\alpha_1$ radiation ($\lambda$ = 1.5406~\AA). Nelson-Riley fits \cite{Nelson1945_extrapolation} were used to calculated $c$-axis lattice parameters. 
Reciprocal space maps were collected with a PIXcel3D area detector on the same system.

\subsection*{Synchrotron X-ray - Crystal truncation rods}
Crystal truncation rod (CTR) measurements were collected at Beamlines 7ID-C and 33-BM of the Advanced Photon Source using a six-circle Huber diffractometer configured in Psi-Circle geometry in both beamlines. The incident X-ray beam, fine-tuned to an energy of 17.5~keV ($\lambda$ = 0.70846~\AA) using a Si (111) double-crystal monochromator ($\delta$E/E $\approx$ 1~×~10$^{-4}$), was focused to a 30~$\mu$m (vertical) × 50~$\mu$m (horizontal) beam spot using Kirkpatrick–Baez mirrors at 7-ID-C, delivering a total photon flux of $\approx$ 3~×~10$^{12}$ photons/s. The X-ray energy used at 33-BM is 20 KeV ($\lambda$~=~0.6199~\AA)  with a total photon flux of  $\approx$ 6~×~10$^{11}$ photons/s. CTRs, also known as Bragg rod measurements, including both specular (0 0 $L$) and off-specular ($H K L$) reflections, were acquired using an Eiger2 X 500K area detector.
For specular CTR measurements, incident and exit angles were kept equal, while off-specular CTR measurements employed a fixed incident angle of 5\textdegree to maintain the same X-ray footprint. Raw 2D detector images were corrected for background scattering, geometric factors, and non-uniform pixel response. See \ref{sec:supp_chess}A for additional processing details.

\subsection*{Synchrotron X-ray - High dynamic-range reciprocal space mapping} 
High Dynamic Range Reciprocal Space Mapping (HDRM) measurements were conducted at the ID4B (QM2 – Q Mapping for Quantum Materials) Beamline at the Cornell High Energy Synchrotron Source (CHESS). Samples were mounted with GE varnish to a sample post such that the film normal was oriented along the $\phi$ axis of a four-circle Huber diffractometer in reflection geometry. All scans were taken at grazing incidence as the sample was continuously rotated around $\phi$ through 0\textdegree\ to 360\textdegree\ in 0.1\textdegree\ steps with 0.1 s counting time per image. Geometric parameters for the large area detector (Pilatus 6M), were determined using standard CeO$_2$ powder data. This measurement was repeated three times per sample, varying the angles $\theta$ (0.5\textdegree, 0.7\textdegree, 1.0\textdegree) and $\chi$ (90\textdegree, 88\textdegree, 95\textdegree). For low-temperature measurements, samples were cooled using a stream of cold helium gas pointed to the sample. Samples on NGO and LAO were measured with an incident X-ray energy of 20 keV, while samples on Sr-containing substrates (STO, SLAO) were measured with an incident X-ray energy of 15 keV to avoid contributions from Sr fluorescence. Data was analyzed with the NXRefine \cite{nxrefine_docs} and NXS-Analysis-Tools \cite{Gomez2026_nxsTools} software packages; see \ref{sec:supp_chess}B-D for additional processing details.


\section*{Author contributions}
The project was conceived by A.Y.J., M.B.-S., A.S.B, and J.A.M.
Thin film synthesis and in-lab X-ray characterization were done by M.B.-S. and A.Y.J., with assistance from D.F.S., G.A.P., J.H., C.M.B., and J.A.M. 
Electrical transport measurements were conducted by M.B.-S and A.Y.J.
Electron microscopy measurements were conducted by L.B. with input from B.H.G. and D.A.M.
Synchrotron X-ray diffraction measurements were performed and analyzed by K.Y.B, D.F.S., D.A.W., and H.Z. at the Advanced Photon Source, and by A.Y.J., M.B.-S., A.B.T., and S.S. at the Cornell High Energy Synchrotron Source.
Structural relaxations by DFT were performed by Y.-F.Z. and A.S.B. 
The manuscript was written by M.B.-S., A.Y.J., and J.A.M. with input from all authors. 

\section*{Materials and correspondence}
Correspondence and material requests should be addressed to Julia A. Mundy at mundy@fas.harvard.edu.

\section*{Competing interests}
The authors declare no competing interests. 

\begin{acknowledgments}

We thank Steven J. Gomez Alvarado for timely support in reciprocal space; Jennifer E. Hoffman, Suzanne Smith, and Jarad A. Mason for helpful discussions and manuscript feedback; Gabriella Foulkes, Avinashi Bhandari, Ty Kelliher, and Michelle Zhang for substrates support; Megan Goh for help with X-ray diffraction measurements; and Jim MacArthur for electronics support.

This project was primarily supported by the National Science Foundation under collaborative awards DMR-2323970 and DMR-2323971.
This research was performed on APS beam time award(s) (DOI: \url{https://doi.org/10.46936/APS188499/60013161}) from the Advanced Photon Source, a U.S. Department of Energy (DOE) Office of Science user facility operated for the DOE Office of Science by Argonne National Laboratory under Contract No. DE-AC02-06CH11357.
This work made use of the Cornell Center for Materials Research shared instrumentation facility. The Thermo Fisher Spectra 300 X-CFEG was acquired with support from PARADIM, an NSF-MIP (DMR-2039380), and Cornell University.
This work was performed in part at the Harvard University Center for Nanoscale Systems (CNS); a member of the National Nanotechnology Coordinated Infrastructure Network (NNCI), which is supported by the National Science Foundation under NSF award no. ECCS-2025158.
This work is based on research conducted at the Center for High-Energy X-ray Sciences (CHEXS), which is supported by the National Science Foundation (BIO, ENG and MPS Directorates) under award DMR-2342336.

D.F.S., G.A.P., and J.A.M. acknowledge support from US Department of Energy, Office of Basic Energy Sciences, Division of Materials Sciences and Engineering, under award no. DE-SC0021925. 
A.Y.J. and D.F.S. acknowledge support from the NSF Graduate Research Fellowship No. DGE-1745303, and M.B.-S. from No. DGE-2141064.
A.Y.J. was also supported by the Paul and Daisy Soros Fellowship for New Americans and the Ford Foundation.  J.A.M. acknowledges support from a Packard Fellowship. 
L.B. and D.A.M. acknowledge support by the NSF Platform for the Accelerated Realization, Analysis, and Discovery of Interface Materials (PARADIM) under cooperative agreement No. DMR-2039380. 
B.H.G. was supported by the Max Planck Society
A.S.B and Y.-F.Z. acknowledge support from NSF grant no. DMR-2323971 and the ASU Research Computing Center for high-performance computing resources. 
\end{acknowledgments}

\newpage
\clearpage
\onecolumngrid

\setcounter{figure}{0}
\setcounter{table}{0}

\makeatletter
\renewcommand{\theequation}{ED\arabic{equation}}
\renewcommand{\thefigure}{ED\arabic{figure}}
\renewcommand{\thetable}{ED\arabic{table}}

\renewcommand{\fnum@figure}{\textbf{Extended Data Fig.~\thefigure}}

\makeatother


\clearpage
\newpage
\centering
\textbf{\large References}
\putbib[comboBib]
\end{bibunit}

\begin{bibunit}[apsrev4-2-titles]
\defaultbibliographystyle{apsrev4-2-titles}
\defaultbibliography{supplement}

\newpage
\clearpage
\onecolumngrid
\begin{center}


\textbf{\large Supplemental Information:}

\textbf{\large \mytitle} 
\end{center}

\setcounter{equation}{0}
\setcounter{figure}{0}
\setcounter{table}{0}
\setcounter{page}{1}
\setcounter{section}{0}
\setcounter{subsection}{0}

\renewcommand{\thesection}{}   
\renewcommand{\thesubsection}{Supplementary Note \arabic{subsection}}

\makeatletter
\renewcommand{\theequation}{S\arabic{equation}}
\renewcommand{\thefigure}{S\arabic{figure}}
\renewcommand{\thetable}{S\arabic{table}}
\renewcommand{\bibnumfmt}[1]{[S#1]}

\subsection{Synthesis and basic structural characterization}
\label{sec:supp_growth}

\begin{table}[h]
\caption{\textbf{\Lafour experimental structural parameters.} Substrate lattice parameters retrieved from supplier values; nominal strain calculated from pseudo-cubic and pseudo-tetragonal lattice parameters as following ($(a_{sub, pc} - a_{bulk, pt}) / a_{bulk,pt}$ where $a_{bulk, pt} = 3.845$ \AA \cite{Ling2000_RP123neutron}; film $c$ lattice parameters and errors fit from Nelson-Riley fits to $\theta - 2\theta$ scans shown in Fig.~\ref{fig:ext_labXRD}.
}
\label{tab:supp_structparams}
\centering
\begin{tabular}{lcccc}
\hline
 & SrTiO$_3$ & NdGaO$_3$ & LaAlO$_3$ & SrLaAlO$_4$ \\
\hline
pseudo-cubic $a,b$ axis (\AA)& 3.905 & 3.858 & 3.787 & 3.756 \\

nominal strain (\%) 
& 1.56 & 0.33 & -1.51 & -2.32 \\

film $c$ lattice axis (\AA)& $27.69 \pm 0.08$ 
& $27.89 \pm 0.07$& $28.22 \pm 0.27$ 
& $28.72 \pm 0.44$ \\

orthorhombic unit cell volume (\AA$^3$)& 844.5 & 830.1 & 809.5 & 810.3 \\

sample film thickness (RP units) 
& 6 & 6 & 6 & 4 \\

sample film thickness (nm) 
& 8.31 & 8.37 & 8.47 & 5.74 \\

\hline
\end{tabular}
\end{table}

\begin{figure*}[!htb]
    \includegraphics[clip=true,width=1.0\columnwidth]{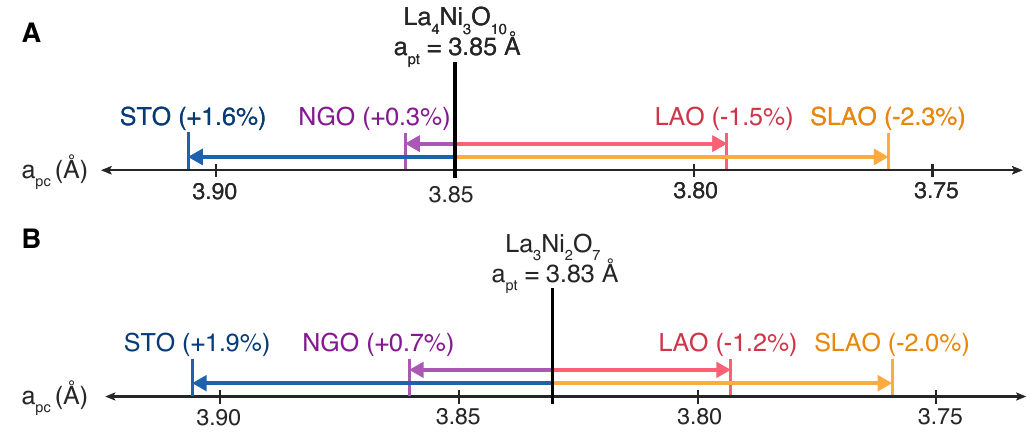}
    \caption{
    \textbf{Comparison of RP nickelate strain states on various pseudo-cubic substrates.} \textbf{a)} \LNO ($n=2$). \textbf{b)} \Lafour ($n=3$). Superconductivity is observed in \LNO on SLAO \cite{Ko2024_La327ambient} but not on \Lafour. Bulk nickelate lattice constants taken from \cite{Wang2025_La3Ni2O7_structure} and \cite{Ling2000_RP123neutron}, respectively.}
    \label{fig:supp_straincomp}
\end{figure*}

\begin{figure*}[!htb]
    \includegraphics[clip=true,width=0.95\columnwidth]{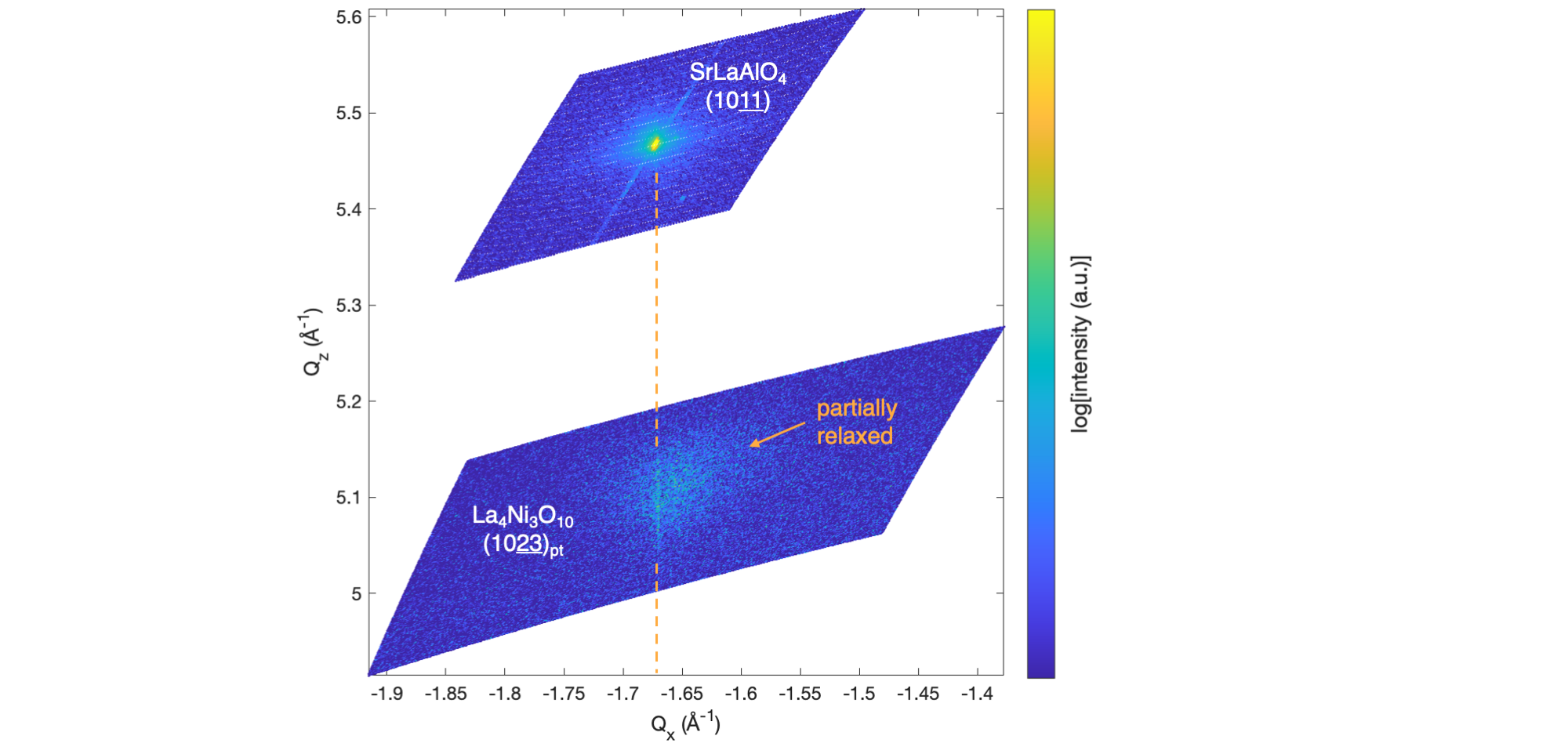}
    \caption{
    \textbf{RSM of thicker \Lafour film (6 RP) on SLAO.} The diffuse intensity to the side of expected strained Q$_x$ demonstrates partial strain relaxation beyond a critical film thickness, as referenced by the arrow.
    }
    \label{fig:supp_RSM_relaxation}
\end{figure*}

\begin{figure*}[!htb]
    \includegraphics[clip=true,width=0.9\columnwidth]{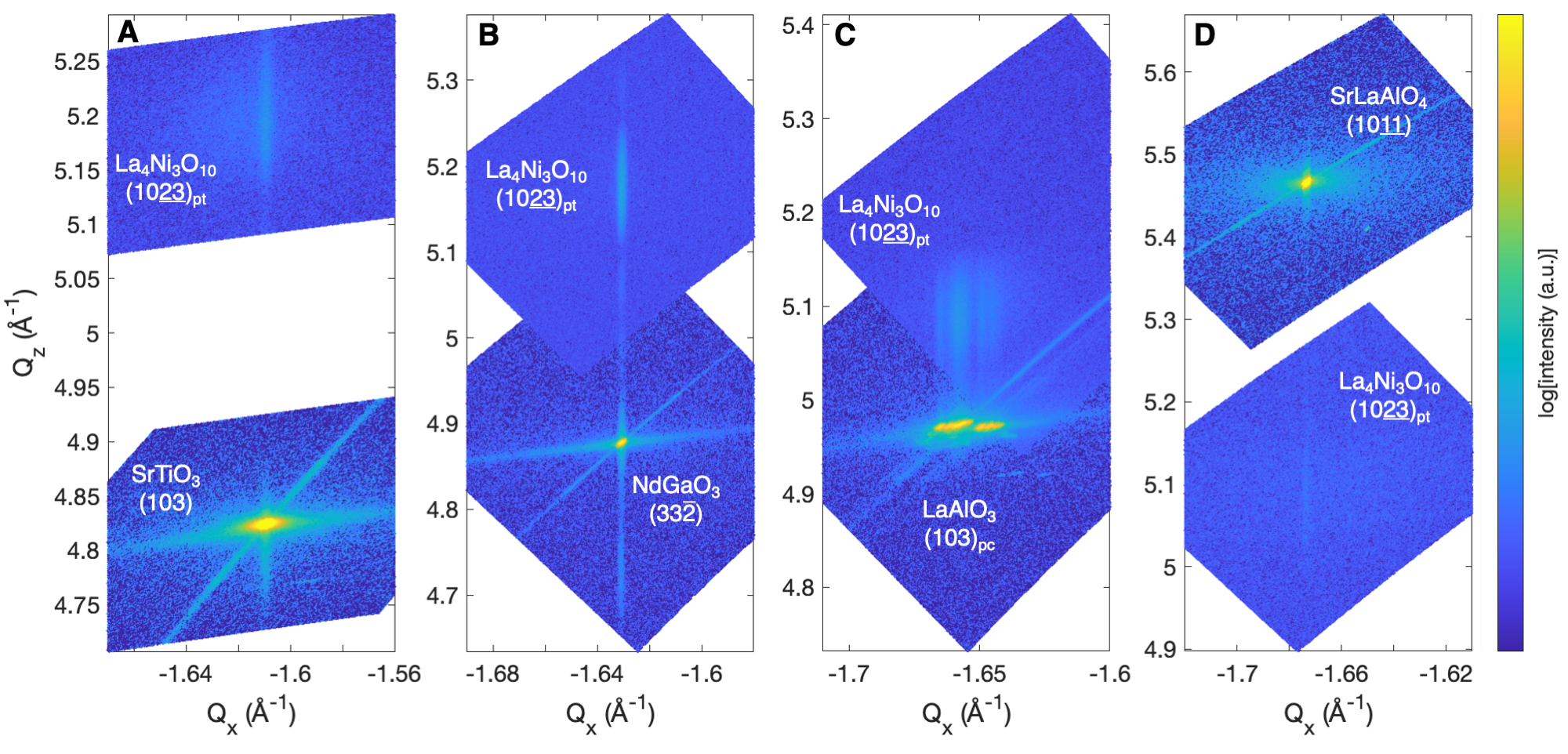}
    \caption{
    \textbf{RSMs of the \Lafour strain series.} \Lafour on \textbf{a)} STO, \textbf{b)} NGO, \textbf{c)} LAO, and \textbf{d)} SLAO, where pc, pt subscripts indicate pseudo-cubic, pseudo-tetragonal indexing respectively. Samples on STO, NGO, and LAO are 6 RP units ($\sim$8 nm) thick, while the sample on SLAO is 4 RP units ($\sim$5 nm) thick.}
    \label{fig:ext_RSM}
\end{figure*}

\begin{figure*}[!htb]
    \includegraphics[clip=true,width=0.75\columnwidth]{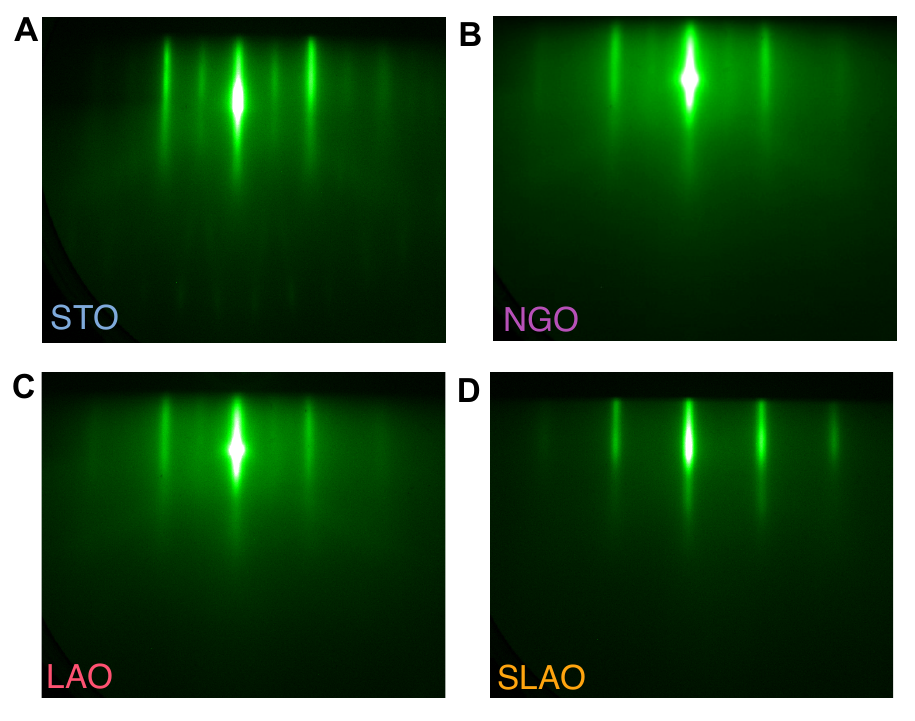}
    \caption{
    \textbf{Post-synthesis RHEED images along the pseudo-cubic [110]$_t$ direction}. \Lafour on \textbf{a)} STO, \textbf{b)} NGO, \textbf{c)} LAO, and \textbf{d)} SLAO substrates. Clear streaks in the RHEED patterns without evidence of secondary phases indicate high quality of films.}
     \label{fig:supp_RHEED}
\end{figure*}

\begin{figure*}[!htb]
    \centering
    \includegraphics[width=0.75\linewidth]{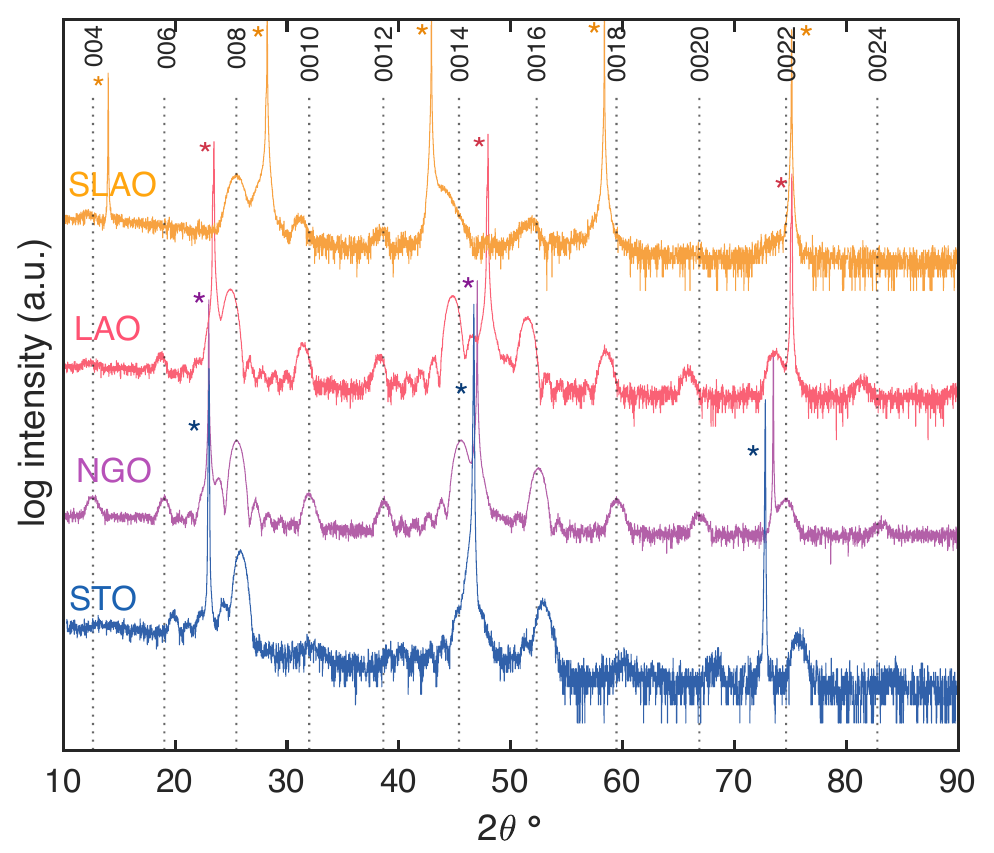}
    \caption{
    \textbf{Lab-based $\theta - 2\theta$ XRD of the \Lafour strain series.} 
    Asterisks denote substrate peaks; dashed lines denote bulk \Lafour peak locations. Synchrotron XRD of the same (00$L$) scans are presented in Fig. \ref{fig:supp_CTR00L}.
    }
    \label{fig:ext_labXRD}
\end{figure*}

\begin{figure*}[!htb]
    \includegraphics[clip=true,width=1.0\columnwidth]{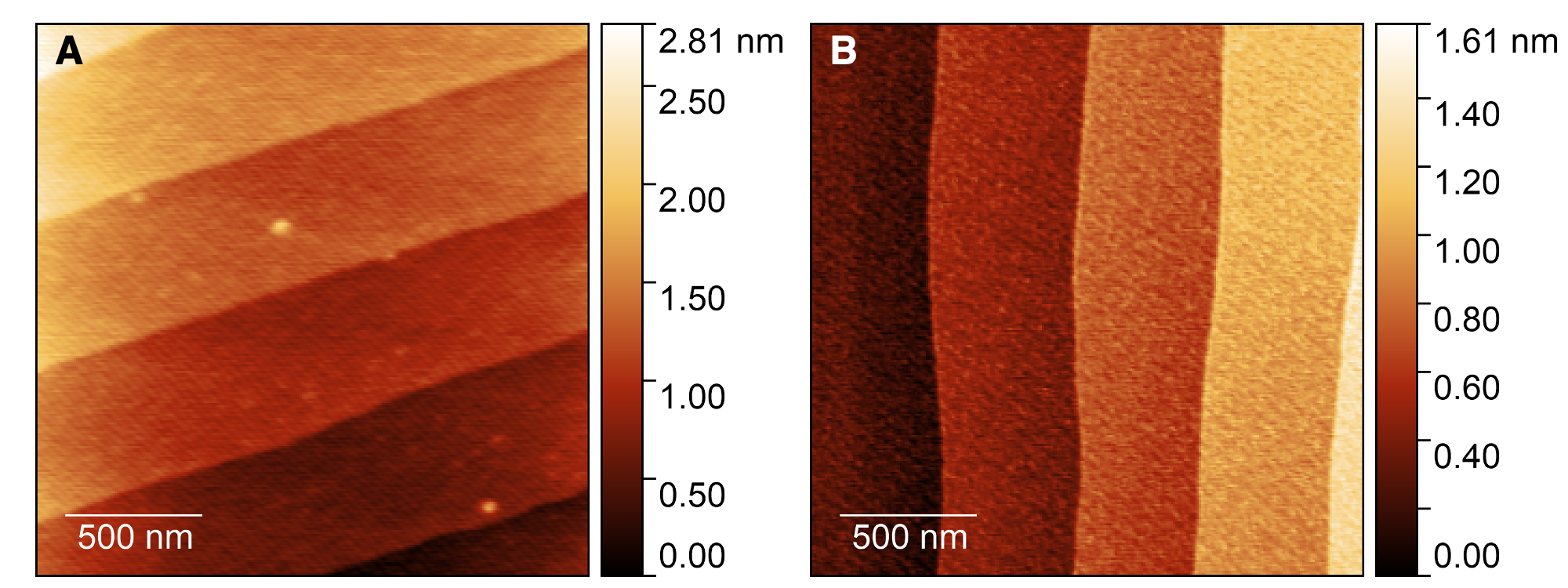}
    \caption{
    \textbf{Atomic force microscopy (AFM) of \Lafour sample.} a) AFM of prepared NGO substrate before MBE synthesis, and b) AFM of \Lafour thin film synthesized on the same NGO substrate. The underlying step terrace structure of the substrate is preserved, 
    indicating exceptionally smooth films.
}
    \label{fig:supp_AFM}
\end{figure*}

\clearpage
\newpage
\subsection{Transport characterization}
\label{sec:supp_transport}

\noindent \underline{A. Resistivity measurements}

\begin{figure*}[!htb]
    \includegraphics[clip=true,width=\columnwidth]{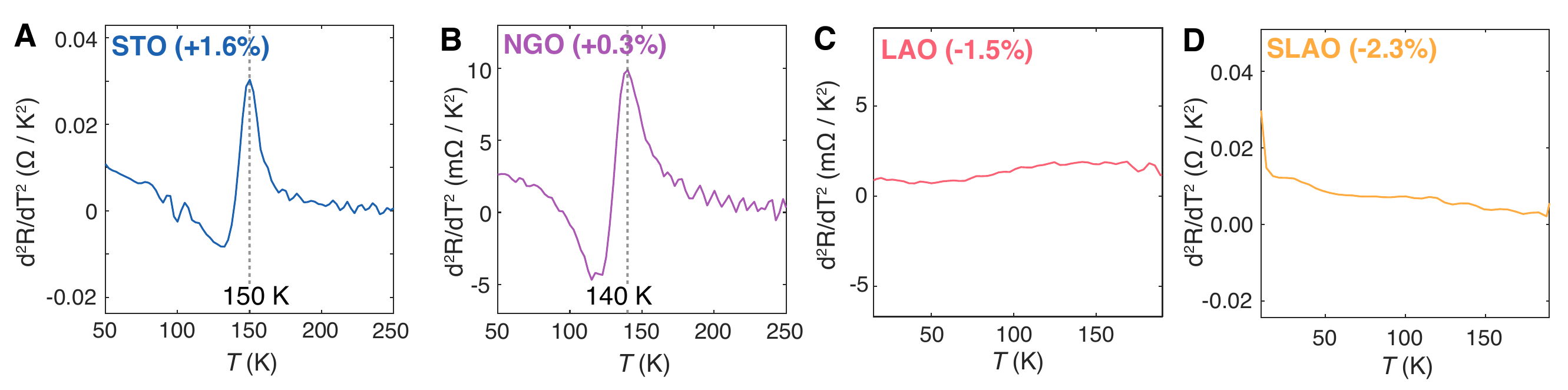}
    \caption{\textbf{Second derivatives of the samples in Fig.~\ref{fig:transport} of the main text, from which \Tdw is determined}. 
    \textbf{a)} STO.
    \textbf{b)} NGO.
    \textbf{c)} LAO.
    \textbf{d)} SLAO. 
    A peak from the CDW transition is present under tensile strain and not detected under compressive strain.
    }
     \label{fig:supp_mainFig_derivs}
\end{figure*}

We define \Tdw as the point of maximum change in slope, which is the peak in the second derivative of the resistance vs. temperature as shown in Fig.~\ref{fig:supp_mainFig_derivs}. This peak is present for for both \Lafour tensile strain states (NGO, STO), with a moderately enhanced temperature on STO. Under compressive strain (LAO, SLAO), there is no visible peak in the first or second derivative, which is consistent with absence of a visible metal-to-metal transition in the resistance vs. temperature measurement. 

\begin{figure*}[!htb]
    \includegraphics[clip=true,width=\columnwidth]{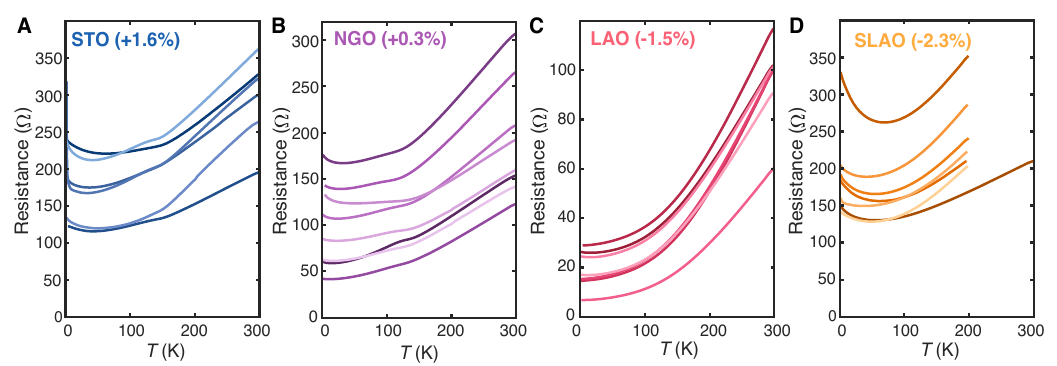}
    \caption{\textbf{Resistance vs. temperature of several \Lafour samples across the strain series.}
    Data is shown for \Lafour on \textbf{a)} STO ($+1.6\%$).
    \textbf{b)} NGO ($+0.3\%$).
    \textbf{c)} LAO ($-1.5\%$).
    \textbf{d)} SLAO ($-2.3\%$). Some samples on SLAO are not measured above 200 K to minimize oxygen loss under vacuum. 
    }
     \label{fig:supp_manyRTs}
\end{figure*}

\begin{figure*}[!htb]
    \includegraphics[clip=true,width=0.6\columnwidth]{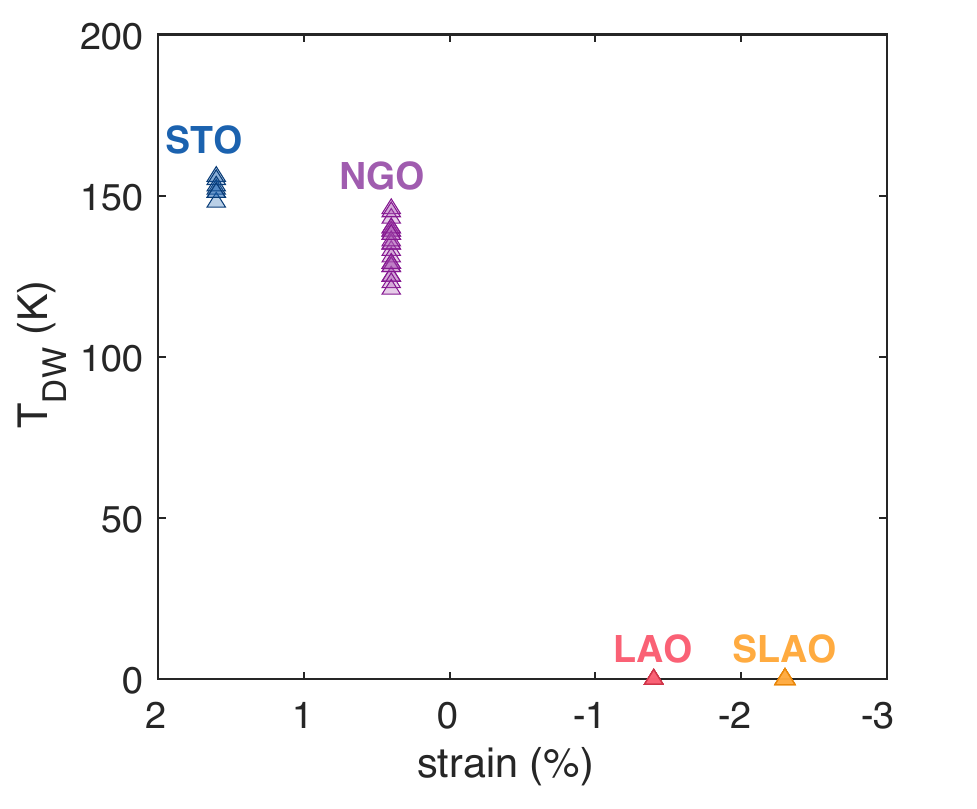}
    \caption{\textbf{Compiled \Tdw from resistance measurements as a function of epitaxial strain}. Each data point is extracted from a single sample; multiple samples are measured on LAO and SLAO as documented in Table \ref{tab:cdw_averages}.
    }
     \label{fig:ext_dome_allpoints}
\end{figure*}

These results are reproduced across several samples grown in different batches spanning several months. We included the results of these efforts in Fig.~\ref{fig:supp_manyRTs}, where \Lafour samples under tensile strain (STO, NGO) consistently show a density wave transition, while those under compressive strain (LAO, SLAO) do not. A compilation plot of the \Tdw from each individual measurement as a function strain is shown in Fig.~\ref{fig:ext_dome_allpoints}. From these results we confirm the systematic enhancement of the transition temperature with increased tensile strain (STO) and absence of density wave under compressive strain; tabulated means and standard deviations across the sample series are included in Table~\ref{tab:cdw_averages}. 


\begin{table}[!htb]
\centering
\caption{
\textbf{\Lafour density wave transition temperature statistics.}}
\label{tab:cdw_averages}
\begin{tabular*}{\textwidth}{@{\extracolsep{\fill}}lccccc}
\hline
Substrate & Strain (\%)  & Mean \Tdw (K) & Std. Dev. (K) & \# samples \\
\hline
STO  & 1.6  & 151.50 & 3.34 & 7\\
NGO  & 0.3  & 133.91 & 7.04 & 22\\
LAO  & -1.5 & none & none & 9\\
SLAO & -2.3 & none & none & 7\\
\hline
\end{tabular*}
\end{table}

\newpage
\noindent \underline{B. Hall measurements}
 
\begin{figure*}[!htb]
    \includegraphics[clip=true,width=0.7\columnwidth]{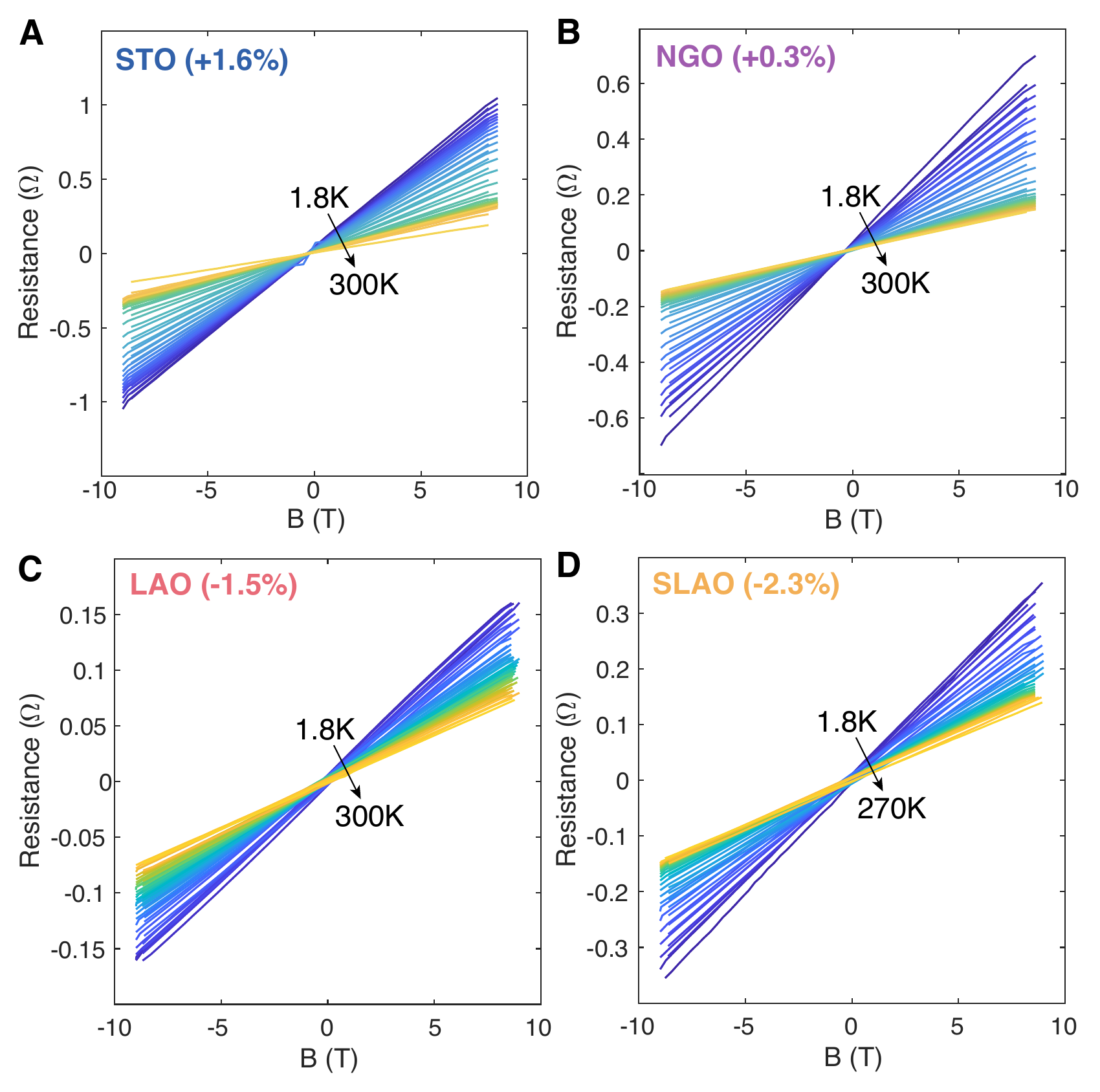}
    \caption{
    \textbf{Antisymmetrized transverse resistance $R_{xy}$ for the samples presented in Fig.~\ref{fig:transport} of the main text.}
    Data is shown for \Lafour on \textbf{a)} STO, \textbf{b)} NGO, \textbf{c)} LAO, and \textbf{d)} SLAO.}
     \label{fig:supp_rxy}
\end{figure*}

\begin{figure*}[!htb]
    \includegraphics[clip=true,width=0.5\columnwidth]{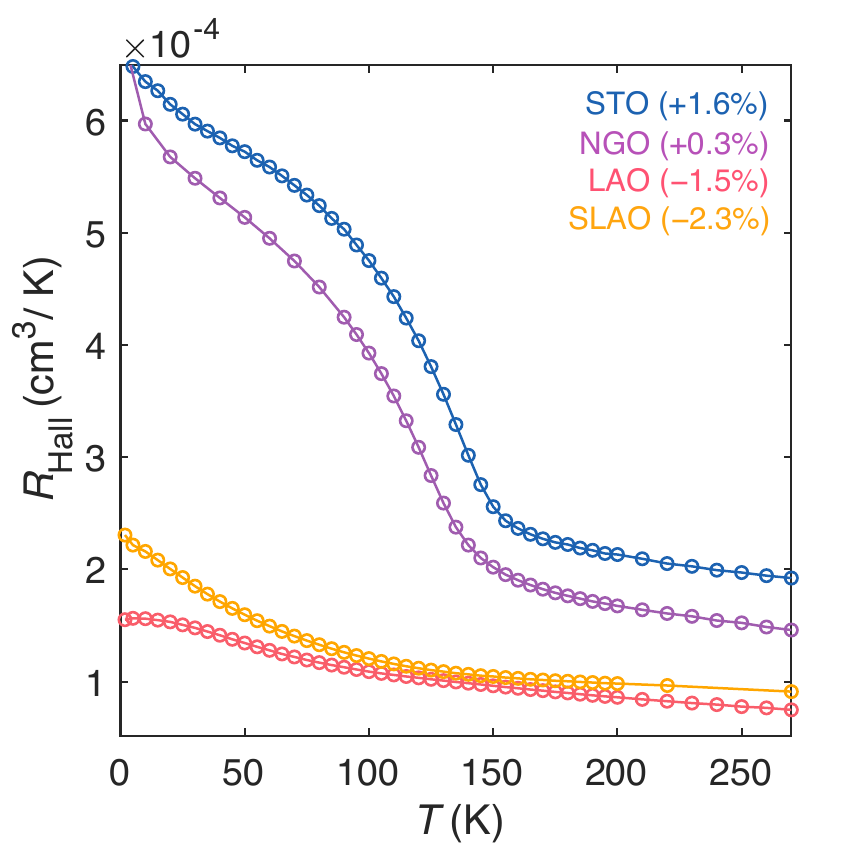}
    \caption{
    \textbf{Hall coefficient $R_{\mathrm{Hall}}$ of the samples presented in Fig.~\ref{fig:transport} of the main text, from which carrier density $n_{\mathrm{Hall}}$ is calculated.
    }
    }
     \label{fig:supp_Rh}
\end{figure*}

To generate the Hall data presented in Fig.~\ref{fig:transport} of the main text, we measure transverse resistance ($R_{xy}$) at a fixed temperature, sweeping field from -9 T to 9 T. For all samples, measurements are taken at base temperature (1.8 K) and in increments of 5 K from 5 K to 200 K. Additional measurements are taken every 10 K from 200 K to 300 K for \Lafour films on STO, NGO, and LAO. For the \Lafour film on SLAO, we took sparser data points above 200 K (220 K and 270 K) to minimize time under vacuum and resultant oxygen loss. We then antisymmetrize $R_{xy}$ at each temperature to reduce geometric contribution from imperfect contacts; the antisymmetrized $R_{xy}$ data for the samples presented in the main text Fig.~\ref{fig:transport}C are shown in Fig.~\ref{fig:supp_rxy}. The fitted slope of the $R_{xy}$ data is the Hall coefficient $R_{\mathrm{Hall}}$ (Fig.~\ref{fig:supp_Rh}), from which we calculate carrier densities $n_{\mathrm{Hall}}$ by the relation $R_{\mathrm{Hall}} = \frac{1}{n_{Hall} ~*~ q}$. 

The sign of $R_H$ corresponds to the type of majority carrier in the system, where a positive coefficient indicates  majority hole-carriers \cite{Dunlap2019_ElectronsInSolids} as observed in our \Lafour films. Interestingly, while there does not seem to be a clear density wave transition in compressively strained \Lafour,  $n_{\text{Hall}}$ and $R_{\text{Hall}}$ still exhibit a nonlinear temperature dependence, potentially suggestive of electronic correlations beyond that expected from a ``normal'' metal. Further characterizations on this intermediate state -- particularly in comparison to the suppressed density wave state found in bulk -- would be an exciting future direction. 




\clearpage
\newpage
\subsection{Magnetic Energies from First-Principles}
\label{sec:DFT_mag}

\begin{figure*}[!htb]
    \includegraphics[clip=true,width=0.85\columnwidth]{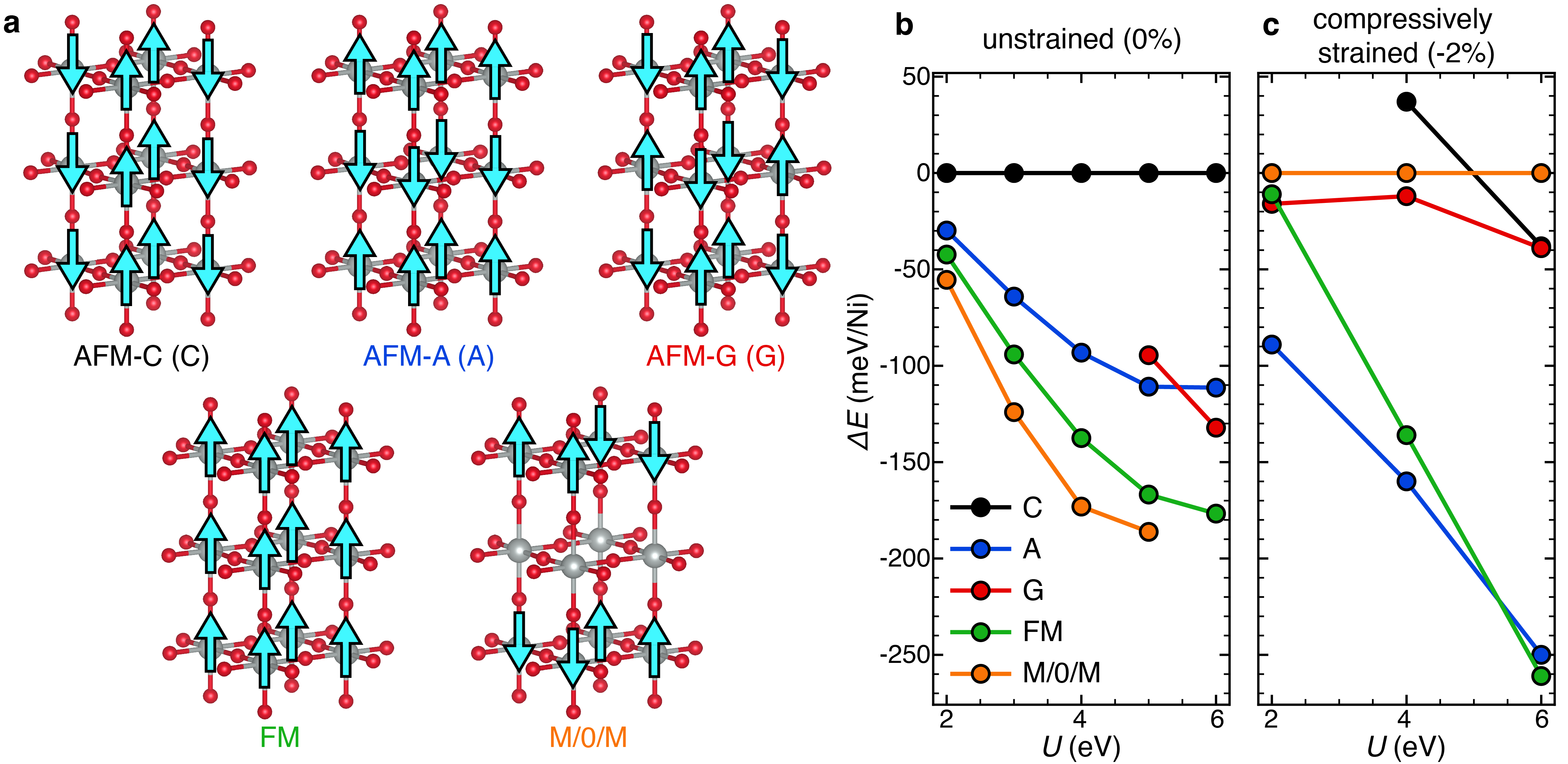}
    \caption{\textbf{Magnetic state energies from first-principles.}
    \textbf{a)} Different spin configurations within the trilayer unit attempted in our first-principles calculations. 
    \textbf{b)} Energy differences as a function of $U$ between different magnetic configurations at unstrained/ambient pressure, and \textbf{c)} under a $-2$\% compressive strain, similar to the experimentally stabilized strain of \Lafour on SLAO ($-2.3$\%).
    }
     \label{fig:en_mag}
\end{figure*}

As mentioned in the main text, at ambient pressure, bulk \Lafour displays metal-to-metal transition that results from intertwined charge and spin density waves. To find the magnetic ground state of \Lafour from first-principles, both in the unstrained limit and under a $-2$\% compressive strain, we performed GGA+$U$ calculations with $U$ values ranging from $2-6$ eV. 
Different magnetic states were targeted, including a ferromagnetic (FM), and antiferromagnetic states: A-type (A-AFM), C-type(C-AFM), and G-type (G-AFM) as depicted in Fig. \ref{fig:en_mag}A. We also targeted a M/0/M (magnetic/non-magnetic/magnetic) state that consists of checkerboard AFM order in the outer planes, which are antiferromagnetically coupled along $c$, and no moment on the inner planes. This corresponds to a magnetic pattern of the six planes in the unit cell (three per trilayer) as follows: $\uparrow$, --, $\downarrow$; $\uparrow$, --, $\downarrow$ where -- represents a spinless layer. This M/0/M magnetic state was found by us (and others) to be the ground state from first-principles in the ambient pressure/unstrained case (see Fig. \ref{fig:en_mag}B \cite{LaBollita2024_La4310}, and it agrees with the intensity distribution of the superlattice reflections derived from single crystal neutron diffraction \cite{Zhang2020_intertwinedDW}. Importantly, the M/0/M state is quickly destabilized energetically under compressive strain, as shown in Fig. \ref{fig:en_mag}C. We note that from experiment, the slight incommensurability of the SDW ordering vector in  La$_4$Ni$_3$O$_{10}$ results in an approximately 5-period stripe in the plane~\cite{Zhang2020_intertwinedDW}, which would be very challenging to calculate from first-principles.

\clearpage
\newpage
\subsection{Post-growth tuning of oxygen content}
\label{sec:supp_ozone}

\begin{figure*}[!htb]
    \includegraphics[clip=true,width=0.7\columnwidth]{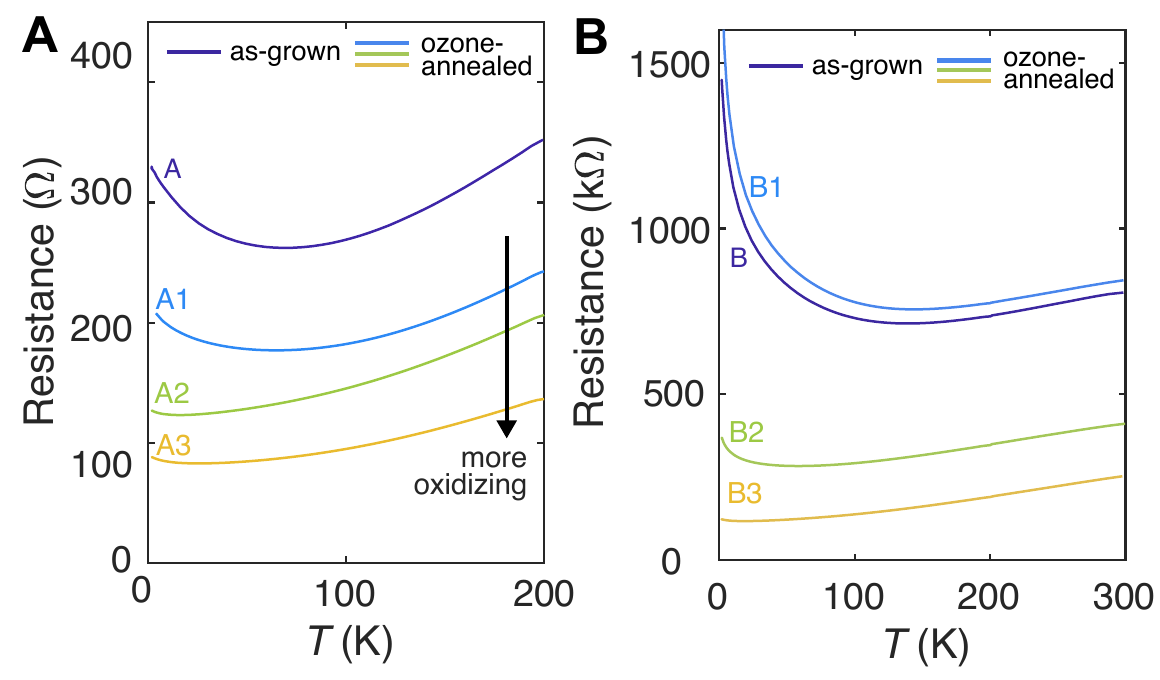}
    \caption{\textbf{Ozone anneals at various temperatures}
    \textbf{a)} Anneals on sample A. A1: 300\textdegree C, 30 min; A2: 300\textdegree C, 1h; A3: 250\textdegree C. 
    \textbf{b)} Anneals on sample B. B1: 300\textdegree C, 30 min; B2: 270\textdegree C, 1h; B3: 250\textdegree C, 1h. 
    }
     \label{fig:supp_ozone_temp}
\end{figure*}
\begin{figure*}[!htb]
    \includegraphics[clip=true,width=0.7\columnwidth]{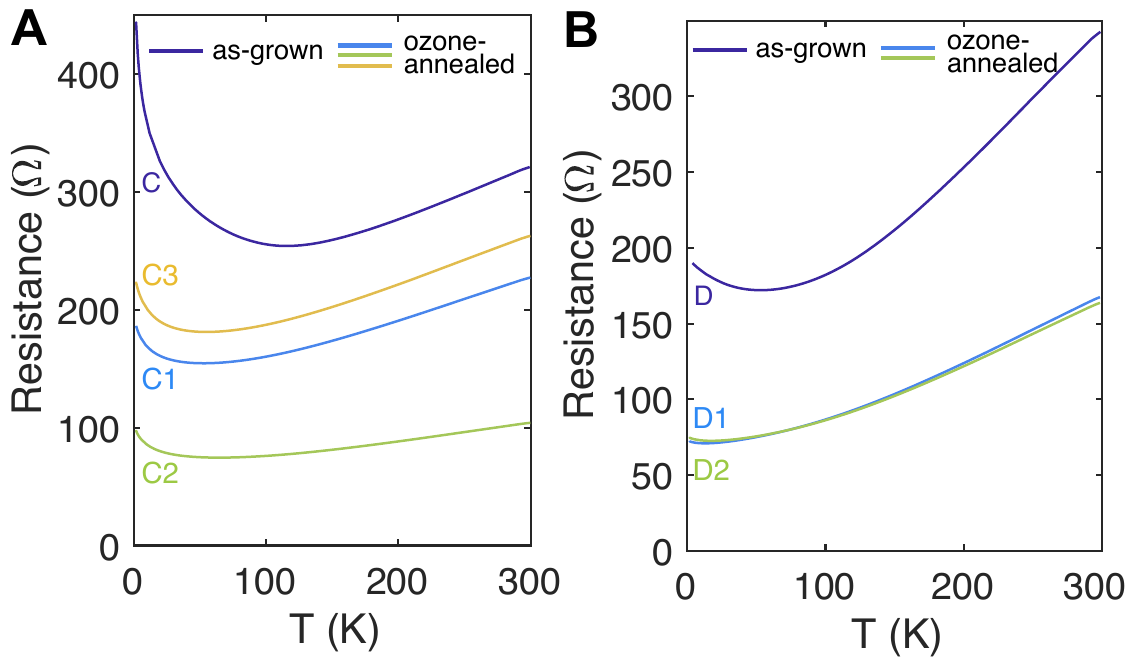}
    \caption{\textbf{Ozone anneals of \Lafour on SLAO with varying length of time.}
    \textbf{a)} Anneals on sample C. C1: 30 min; C2: 1 h; C3: 2 h.
    \textbf{b)} Anneals on sample D. D1: 1 h; D2: 2 h.
    All anneals were done at 300\textdegree C.}
     \label{fig:supp_ozone_time}
\end{figure*}
\begin{figure*}[!htb]
    \includegraphics[clip=true,width=1.0\columnwidth]{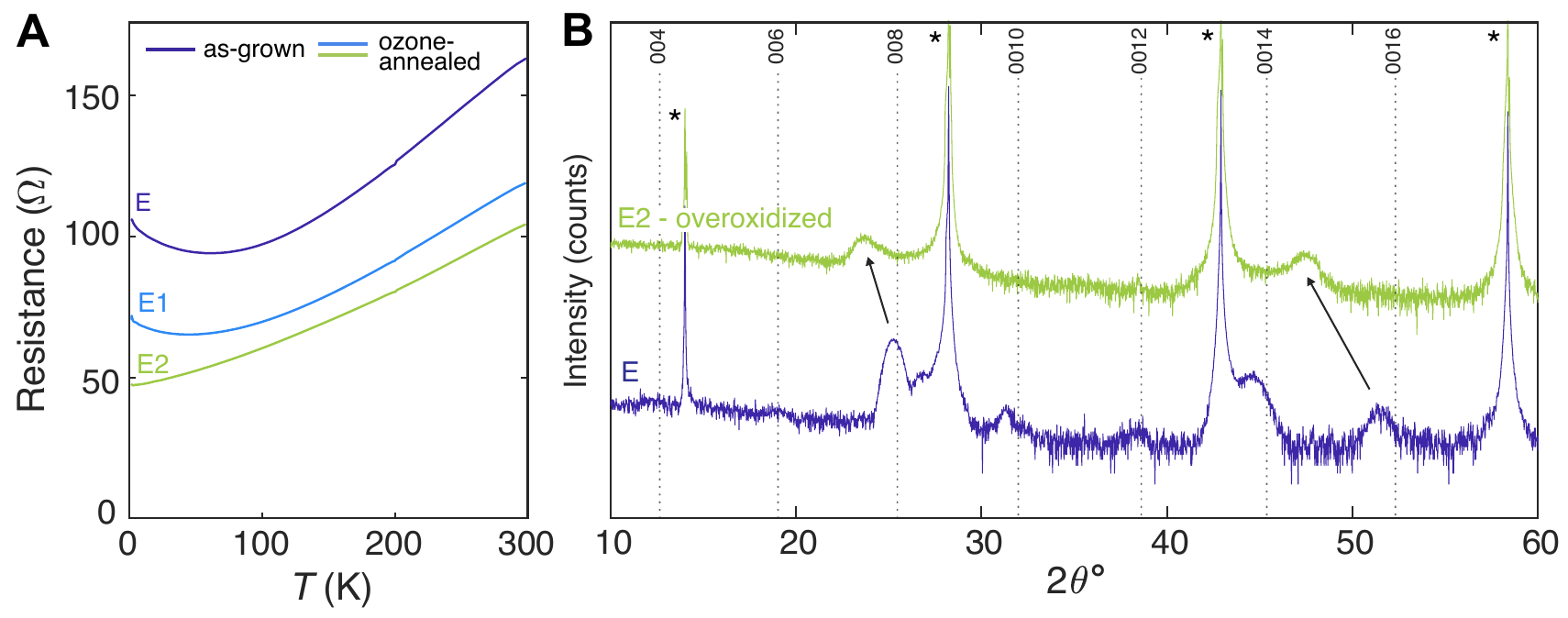}
    \caption{\textbf{Ozone anneals of  \Lafour on SLAO with varying ozone generator power.}
    \textbf{a)} Transport of anneals on sample E, where ozone generator power is tuned by a potentiostat. E1: 1.0 k$\Omega$ power; E2: 1.5 k$\Omega$. 
    \textbf{b)} X-ray diffraction of the higher generator power sample E2, which has transformed to a distinct structural phase upon too high of ozone concentration. 
    }
     \label{fig:supp_ozone_power}
\end{figure*}

While all of the strained \Lafour thin films are susceptible to oxygen loss over time and under vacuum, this effect seems to be exacerbated for samples under high compressive strain (SLAO). For example, when left under mild vacuum (1-10 mTorr) in a PPMS cryostat at room temperature for extended periods of time, samples under low strain (NGO) largely retain their as-grown transport, while sample under high compressive strain (SLAO) become significantly less metallic as has been observed in $n=2$ thin films \cite{Liu2025_La2Pr327transport, Liu2026_halfdome}. Given the observed propensity of \Lafour on SLAO to oxygen loss, we performed a series of anneals under ozone to fill oxygen vacancies and better understand the role of oxygen content on electronic properties. These anneals were done spanning a wide range of conditions and in two different systems, including a commercial UV ozone cleaner (SAMCO) and a home-built ozone annealing furnace based off of a 3 g/h quartz tube generator. To minimize uncertainty regarding sample-to-sample variation, each of the sets of anneals presented in Figs. \ref{fig:supp_ozone_temp}-\ref{fig:supp_ozone_power} represent anneals on various pieces cleaved from a single as-grown wafer. Each individual wafer is given a letter index (A-E), while the number denotes the attempted anneal. 

In general, we find that as the oxidizing power of a given anneal increases -- either by varying temperature, time, or ozone concentration -- increased sample metallicity, either by reducing the room temperature resistance or suppressing the low-temperature resistive upturn although without inducing superconductivity across a range of annealing conditions. For example, progressively lowering the temperature of the anneal from 300\textdegree C to 250\textdegree C increased the metallicity of the samples (Fig.~\ref{fig:supp_ozone_temp}). When varying the time of the anneal, we found that the effect of the anneal seems to saturate at $\sim$ 1 h, with longer anneals producing little to no improvement in metallicity (Fig.~\ref{fig:supp_ozone_time}).   

We also varied power input into the quartz tube ozone generator via a potentiometer, where higher resistance corresponds to higher concentrations of ozone. These results are presented in Fig.~\ref{fig:supp_ozone_power}, where at an intermediate power (1 k$\Omega$ on the potentiometer), the sample becomes more metallic. However, upon increasing power further to 1.5 k$\Omega$, the RP structure becomes destabilized and undergoes a structural transformation to a previously reported oxygen-intercalated structure with excess oxygen beyond the regime of RP stoichiometry \cite{Segedin2026_overox}.

\clearpage
\newpage
\subsection{Additional compressive strain}
\label{sec:supp_YAO}

\begin{figure*}[!htb]
    \includegraphics[clip=true,width=1.0\columnwidth]{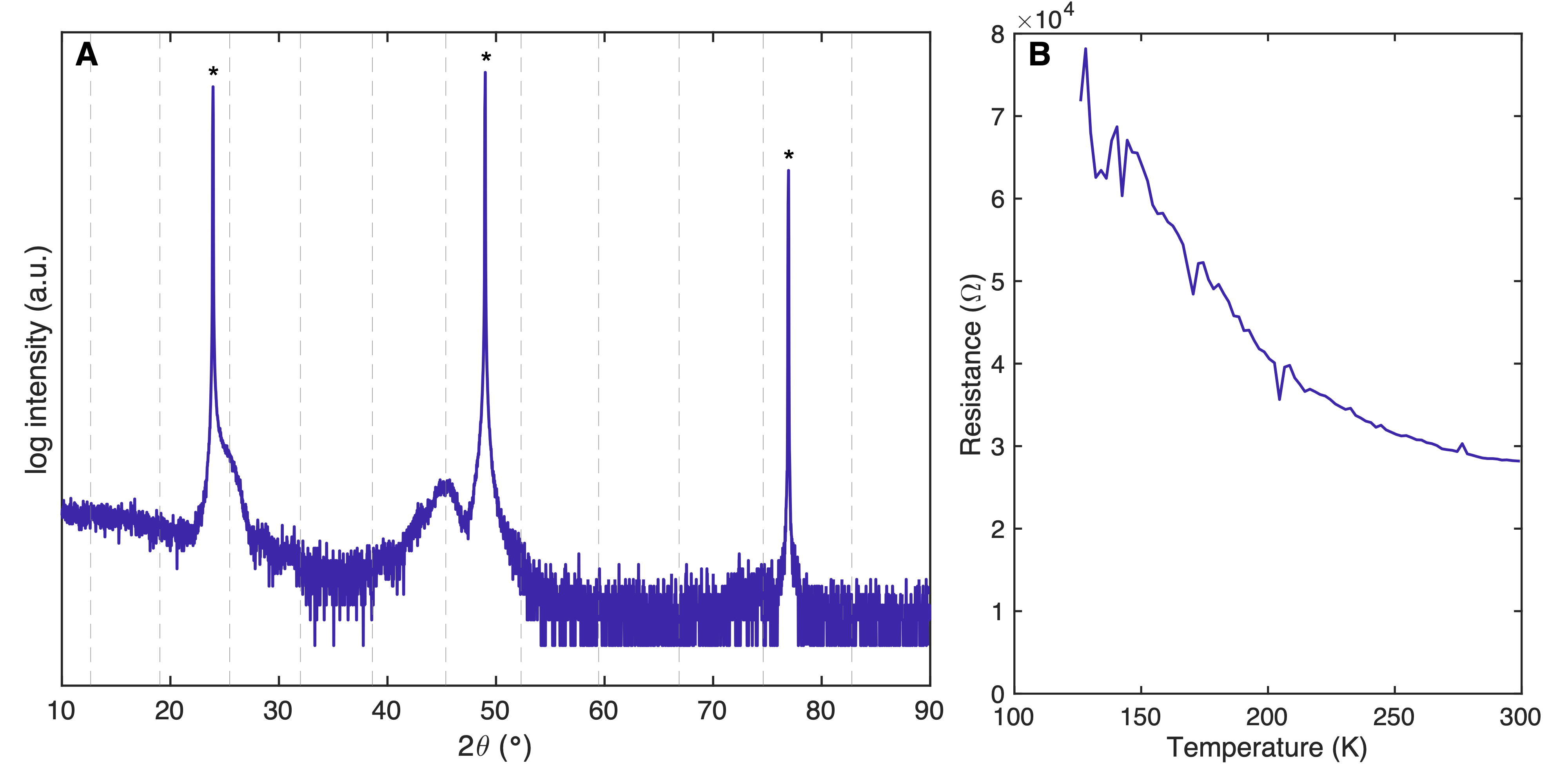}
    \caption{
    \textbf{\Lafour on YAO (101),} which has an in-plane lattice constant of 3.72 Å and imparts a nominal -3.2\% compressive strain on \Lafour.
   \textbf{a)} XRD illustrates the challenges of synthesis on this extreme strain state as shown by significant peak broadness and poor phase purity. Asterisks denote substrate peaks.
   \textbf{b)} Electronic transport exhibits insulating behavior.
}
    \label{fig:supp_YAO}
\end{figure*}

Given the lack of superconductivity in \Lafour on SLAO, we also attempted synthesis of \Lafour under even more compressive strain on the next commercially available substrate, YaAlO$_3$ (101). YAO has a lattice constant of $\sim3.72$ Å, and imparts a nominal $-3.2$\% compressive strain, which is nearly a whole percent more than SLAO. The enormous compressive strain imparted by YAO poses significant synthetic limitations. X-ray diffraction patterns (Fig.~\ref{fig:supp_YAO}A) exhibit broad peaks, and the peaks are ``smeared'' towards the bulk peak locations, indicating relaxation of the film. Likely as a result of this disordered nature, \Lafour films on YAO are also not metallic (Fig.~\ref{fig:supp_YAO}B). 

\clearpage
\newpage
\subsection{Electron microscopy}
\label{sec:supp_STEM}

\begin{figure*}[!htb]
    \includegraphics[clip=true,width=0.75\columnwidth]{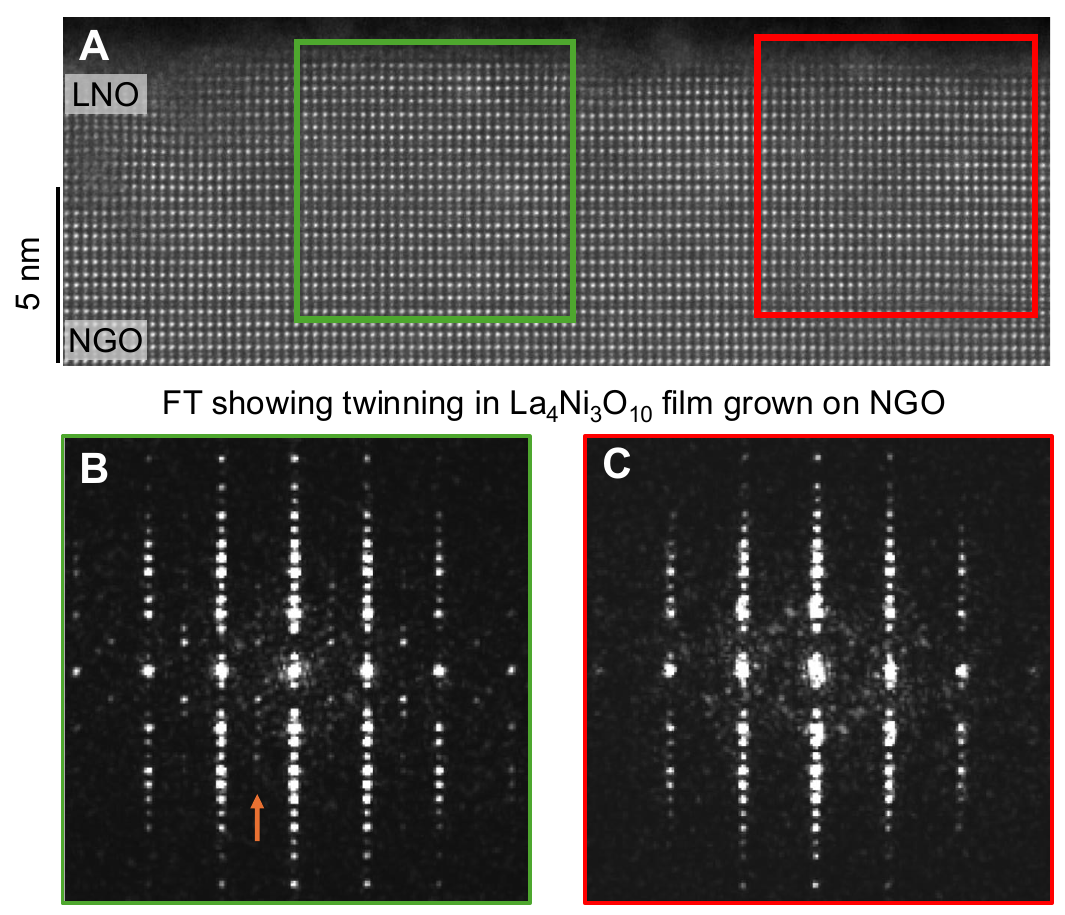}
    \caption{\textbf{Twinning in \Lafour thin films.} 
    \textbf{a)} ADF-STEM image of \Lafour on NGO.
    \textbf{b, c)} Fourier transforms (FTs) of two different areas of the film marked by green and red boxes show distinct peaks. The green region has half-order peaks that are lacking in the red region, which is consistent with the inequivalent \textit{a}, or [100] (green) and \textit{b}, or [010] (red) projections of the $P2_1/a$ structure, indicating the presence of twinning in the film.} 
    \label{fig:supp_NGO_twin}
\end{figure*}

\begin{figure*}[!htb]
    \includegraphics[clip=true,width=0.95\columnwidth]{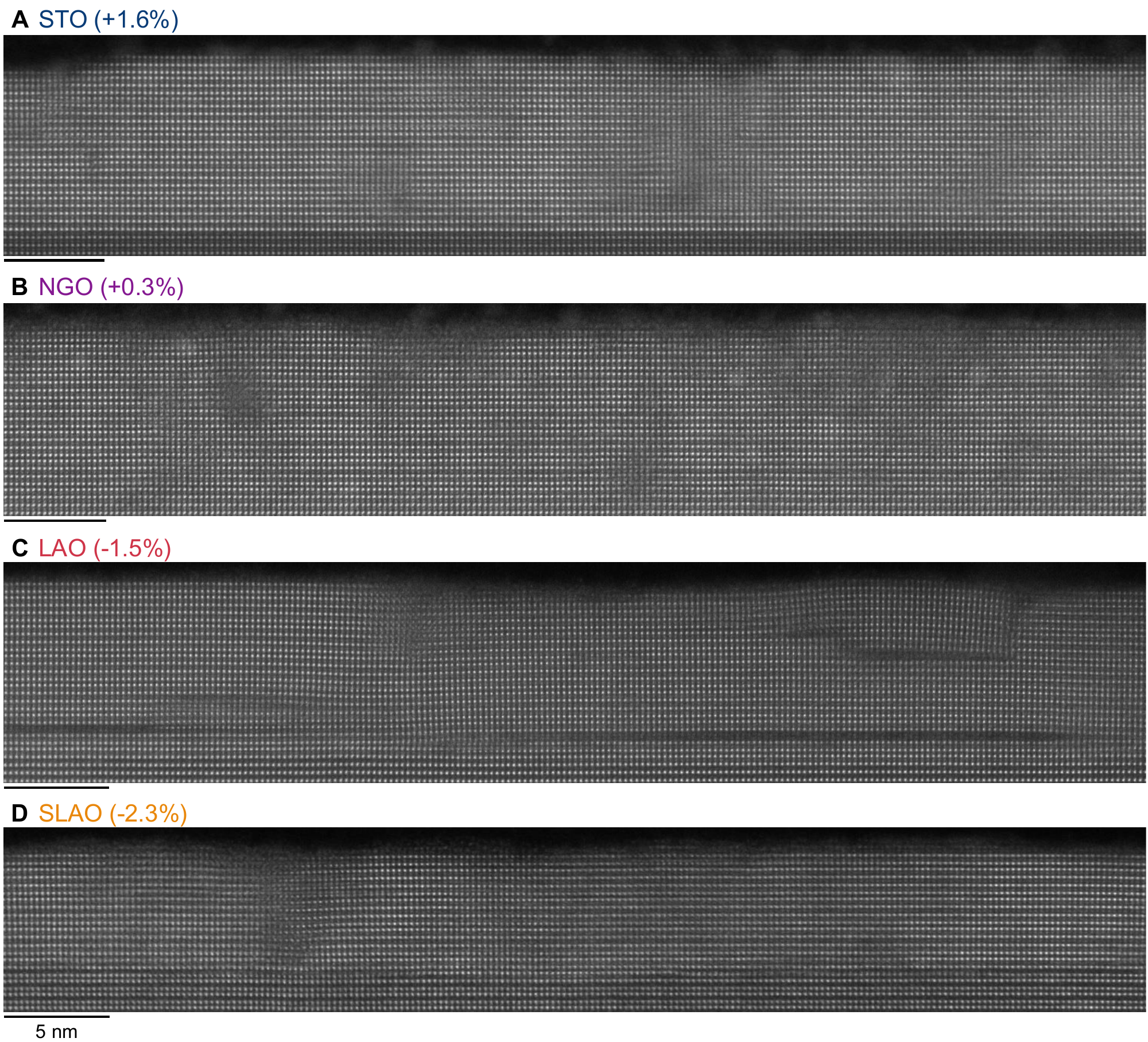}
    \caption{\textbf{Large field-of-view images of \Lafour films. } As synthesized on
    \textbf{a)} STO, \textbf{b)} NGO, \textbf{c)} LAO and \textbf{d)} SLAO. 
    }
    \label{fig:supp_LargeFOV_ADF}
\end{figure*}

For our ADF-STEM measurements, we view \Lafour at room temperature along an orthorhombic cut of the film to observe coherent out-of-plane octahedral rotations that are not visible along the pseudo-cubic projection that is often used to image perovskite-type structures. In a given large field-of-view ADF image (Fig.~\ref{fig:supp_LargeFOV_ADF}), we observe both [100] and [010] orthorhombic-axis projections of \Lafour, indicating that the film is twinned (Fig.~\ref{fig:supp_NGO_twin}) with lateral domain sizes of about 10-20 nm as previously observed in \LNO \cite{Bhatt2025_La327films}. This twin formation occurs due to the lower orthorhombic symmetry of the RP nickelate relative to the pseudo-cubic substrates. 

We note that the symmetry of the bulk of the film behaves independent of the symmetry of the substrate, despite the film being epitaxially strained to the substrate. 
For example, our STEM results indicate that \Lafour hosts out-of-plane octahedral rotations when synthesized on both NGO and STO, despite the fact that NGO hosts rotations of its own GaO$_6$ octahedra, while STO does not host rotations in its TiO$_6$ octahedra.
This observed independence of the \Lafour thin film symmetry from that of the substrate is consistent with other literature on epitaxial transition-metal perovskite oxides, where the symmetry of a substrate extends only a few perovskite unit cells into neighboring higher-symmetry octahedra \cite{Hwang2013_RENiO3_superlattices, Qi2015_LNO_LGO_superlattices}. 


\clearpage
\newpage
\subsection{Structural evolution with strain from first-principles}
\label{sec:dft_struct}

\begin{figure*}[!htb]
    \includegraphics[clip=true,width=0.95\columnwidth]{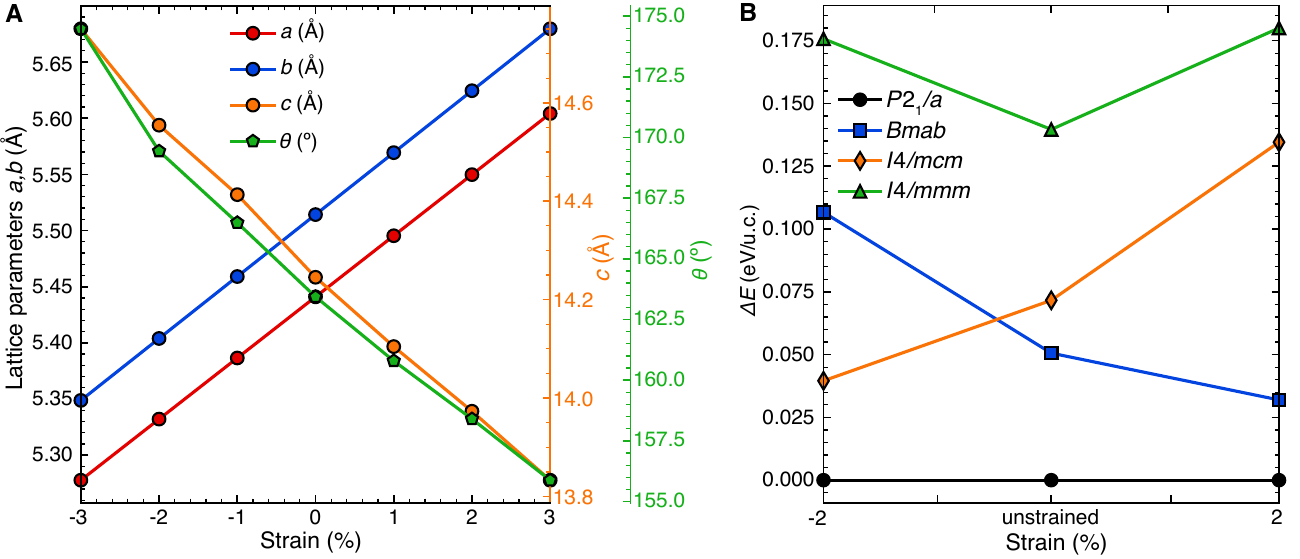}
    \caption{\textbf{Structural evolution with strain from first-principles.}
   \textbf{a)} Lattice constants and vertical Ni-O-Ni bond angle across the apical oxygens with a $P2_1/a$ structure used as a starting unstrained ambient pressure space group symmetry. 
   \textbf{b)} Energy differences between a $P2_1/a$, $Bmab$, $I4/mmm$ and $I4/mcm$ structure for different levels of strain. 
    }
    \label{fig:struct_dft}
\end{figure*}

Fig. \ref{fig:struct_dft} summarizes the structural evolution of La$_4$Ni$_3$O$_{10}$ upon compressive and tensile strain from first-principles calculations. Structural optimizations using $P2{_1}/a$ as the starting space group symmetry clearly show a suppression of the out-of-plane octahedral rotations upon compressive strain, in agreement with experiments, even though 180$^\circ$ Ni-O-Ni bond angles across the apical oxygens are not fully reached in DFT calculations (see Fig. \ref{fig:struct_dft}A).  While from first-principles a $P2{_1}/a$ structure remains the ground state for all strain levels studied here, an $I4/mcm$ structure closely competes in energy in the compressive strain limit (see Fig. \ref{fig:struct_dft}B).

\clearpage
\newpage
\subsection{Multislice electron ptychography}
\label{sec:supp_ptycho}

\begin{figure*}[!htb]
    \includegraphics[clip=true,width=0.95\columnwidth]{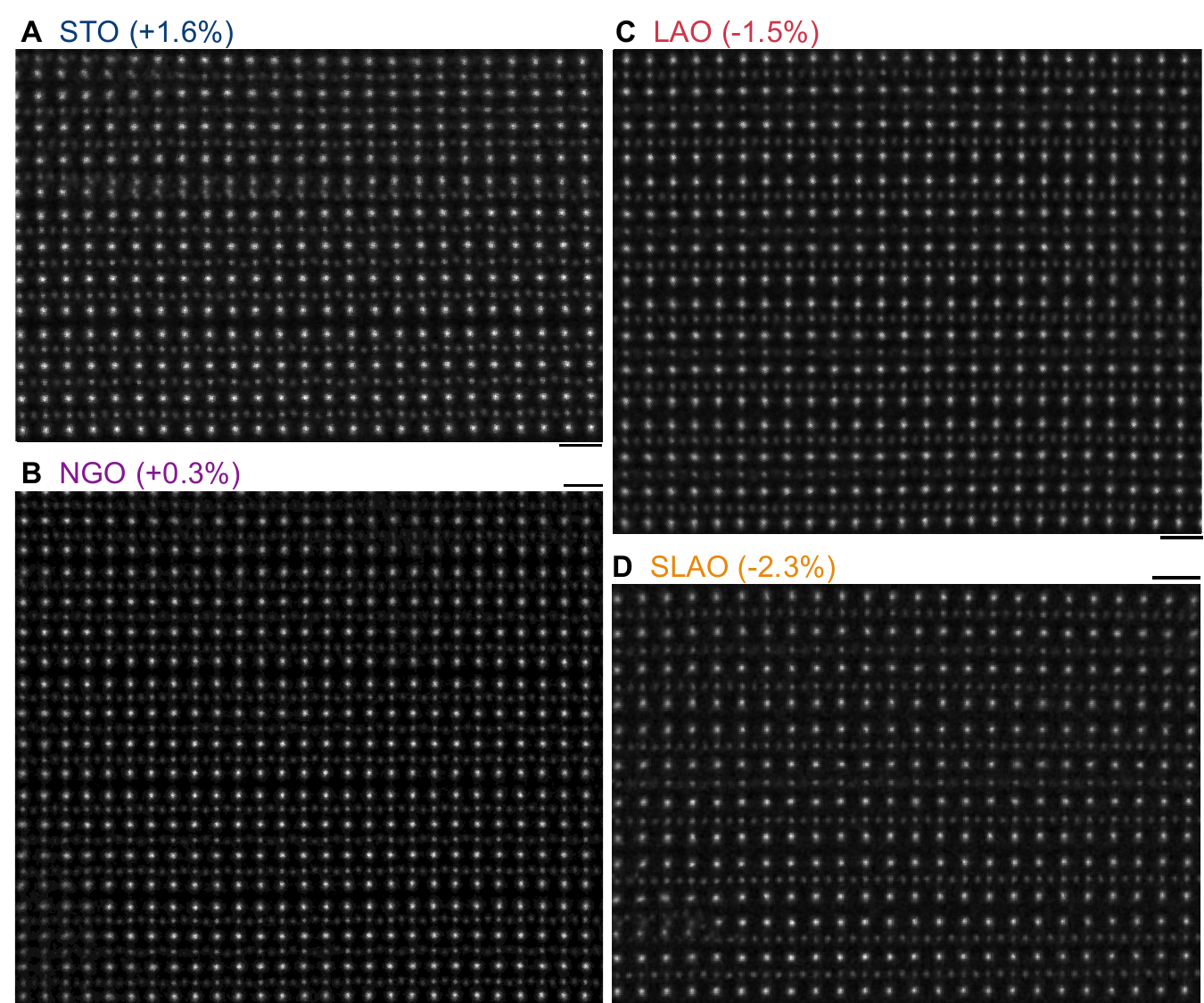}
    \caption{\textbf{Large field-of-view ptychographic reconstruction of \Lafour films presented in Fig.~\ref{fig:xtal_phase} of the main text.} As synthesized on 
    \textbf{a)} STO, \textbf{b)} NGO, \textbf{c)} LAO and \textbf{d)} SLAO. Scale bars are 5 nm.
  }
    \label{fig:supp_LargeFOV_ptycho}
\end{figure*}

\clearpage
\newpage
\subsection{Synchrotron X-Ray diffraction}
\label{sec:supp_chess}

\noindent \underline{A. Crystal truncation rod scattering}

Crystal truncation rod (CTR) scattering is an area-averaged technique that takes advantage of interference between the film and substrate to measure reflections across various places in reciprocal space \cite{Disa2020_CTR,May2010_quantifyRot}. For example, (0 0 $L$) reflection is the specular reflection, which corresponds to typical $\theta-2\theta$ scans collected in lab thin-film X-ray measurements (Fig.~\ref{fig:ext_labXRD}). The geometry of CTR results in clear intensity signals from epitaxial thin films even when measuring samples of $<$10 nm. 

\begin{figure*}[!htb]
    \centering
    \includegraphics[width=0.7\linewidth]{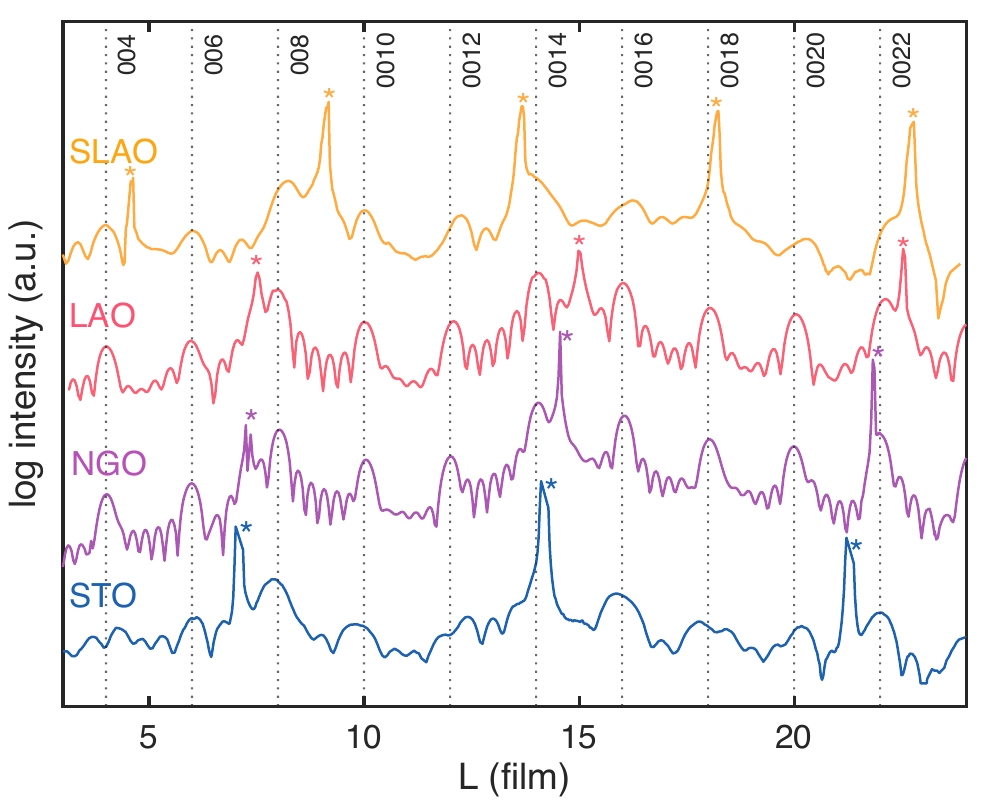}
    \caption{\textbf{Synchrotron CTR (0 0 $L$) scans of \Lafour across the strain series.} $L$ indexed to the \Lafour $c$-axis of each sample; asterisks denote substrate peaks; dashed lines denote peak indexing.}
    \label{fig:supp_CTR00L}
\end{figure*}

For non-specular scans, we base our interpretations based on prior investigations into rotations in perovskites \cite{May2025_bookChapter, May2010_quantifyRot, Yuan2018_octahedralTiltEpitaxy}, particularly for the (1.5 0.5 $L$) CTR which indexes in-plane octahedral rotations. We make this comparison because in the RP structure, octahedra retain the in-plane corner-sharing network equivalent to the traditional perovskite structure within the $a,b$ plane, allowing direct interpretation of in-plane octahedral rotations based on presence (or absence) of diffracted intensity at $L$=integer in (1.5 0.5 $L$). In contrast, the (0.5 0.5 $L$) CTR which indexes out-of-plane octahedral rotations in perovskites cannot be directly interpreted in the RP crystal structure, as the rock salt spacer layers disrupt the corner-sharing octahedral network along the $c$-axis. 

\begin{figure*}[!htb]
    \includegraphics[clip=true,width=0.55\columnwidth]{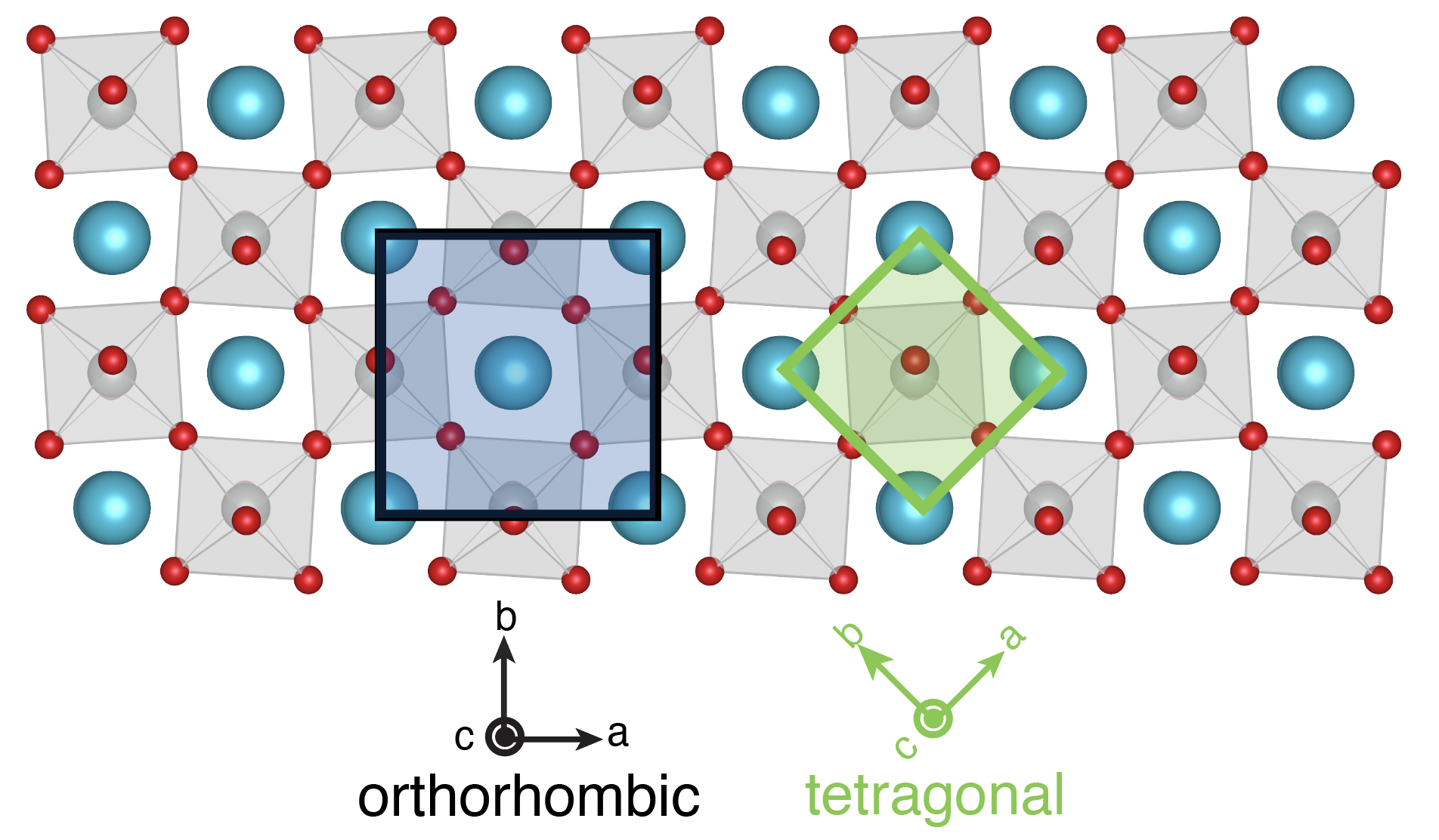}
    \caption{
    \textbf{In-plane view of the orthorhombic and pseudo-tetragonal unit cell definitions for \Lafour.} In the orthorhombic unit cell, $a \neq b$. However, the two in-plane axes are often approximated as $a=b$ and indexed to a pseudo-tetragonal unit cell for comparison to well-studied perovskites. The $P2_1/a$ structure from \cite{Zhang2020_LaPr4310floatingzone} is shown here.
    }
    \label{fig:supp_ortho_vs_tet}
\end{figure*}

All CTR data shown in this manuscript is indexed to the pseudo-tetragonal unit cell (Fig.~\ref{fig:supp_ortho_vs_tet}) to match traditional perovskite indexing. To account for the out-of-plane unit cell difference between the film and the substrate, we rescale $L$ by the ratio of $c_{\text{film}} / c_{\text{substrate}}$, where $c_\text{film}$ is extracted from Nelson-Riley fits \cite{Nelson1945_extrapolation} of the lab-based $\theta$-2$\theta$ scans in Fig.~\ref{fig:ext_labXRD}.



\noindent \underline{B. HDRM Unit cell indexing}

For the HDRM measurements, the 3D dataset is initially refined to the substrate due to the significantly (several orders of magnitude) higher intensity of the substrate relative to the film, which enables automatic peak detection for the substrate symmetry.
With knowledge of the film orientation relative to the substrate, we then recover the \Lafour orientation for each dataset via subsequent rotation and scaling steps \cite{Gomez2026_nxsTools}. We note that the specific series of rotations used recover the film orientation changes between substrate orientations. For example, \Lafour on the (110) plane of NGO requires different rotations compared to \Lafour on STO (100). While bulk \Lafour 
adopts a monoclinic $P2_1/a$ structure, we index our HDRM data to either a pseudo-orthorhombic or pseudo-tetragonal unit cell where $\alpha = \beta = \gamma = 90^\circ$ to facilitate comparison with other characterization methods. The orthorhombic and tetragonal unit cells are rotated 45\textdegree\ relative to each other, with the in-plane lattice parameters of the lower symmetry orthorhombic unit cell being a factor of $\sqrt{2}$ larger in real space (see Fig.~\ref{fig:supp_ortho_vs_tet}). To account for the out-of-plane unit cell difference between the film and the substrate, we rescale $L$ by the ratio of $c_{film} / c_{substrate}$, similar to CTR scans. 

\begin{figure*}[!htb]
    \includegraphics[clip=true,width=0.95\columnwidth]{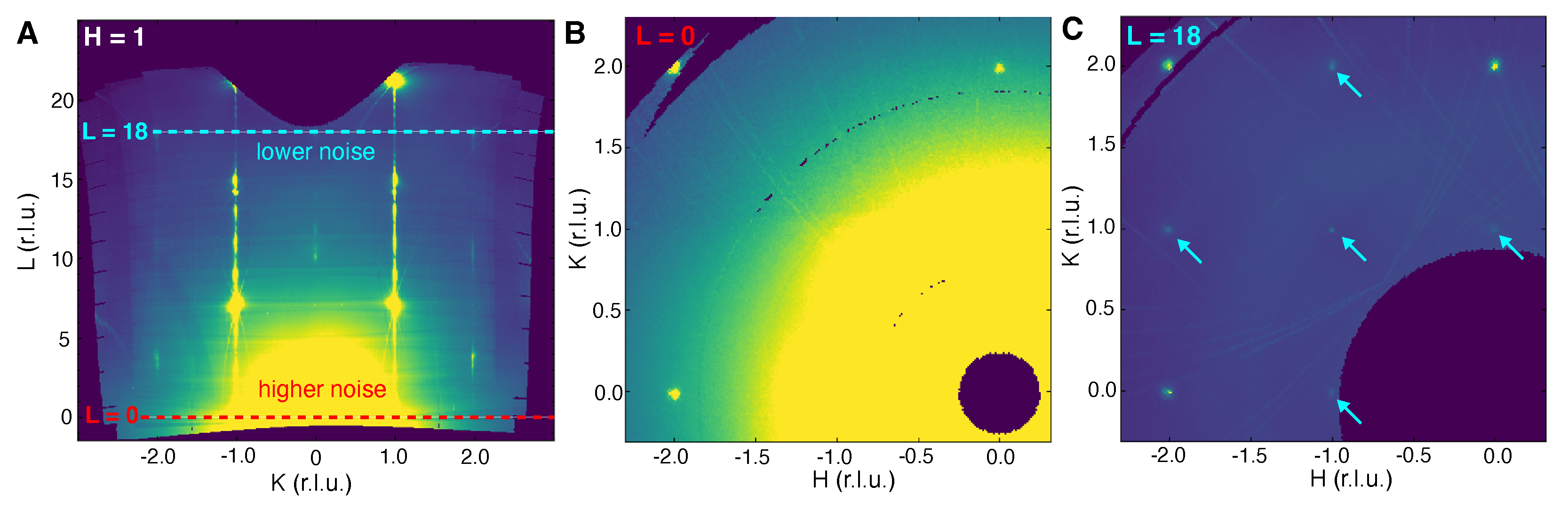}
    \caption{
    \textbf{Detector noise at different $L$ for a representative \Lafour sample on STO.} 
    \textbf{a)} H = 1 cut. \textbf{b)} In-plane cut at $L$ = 0. \textbf{c)} In-plane cut at L=18 showing peaks at (odd odd $L$) which were previously not detected. Data is indexed to the \Lafour orthorhombic unit cell.}
    \label{fig:supp_CHESS_noise}
\end{figure*}

With the 3D dataset re-oriented to the \Lafour film, we cut at arbitrary H, K, or L. To detect the presence (or lack thereof) of weak film peaks such as half-order reflections in H-K planes, we take cuts at relatively high L, where detector noise is minimal as shown in Fig.~\ref{fig:supp_CHESS_noise}A. Features which are otherwise buried below the noise signal become more visible in these cuts at higher $L$ (Fig.~\ref{fig:supp_CHESS_noise}B,C).

\noindent \underline{C. Analysis of room-temperature HDRM measurements}

\begin{figure*}[!htb]
    \includegraphics[clip=true,width=\columnwidth]{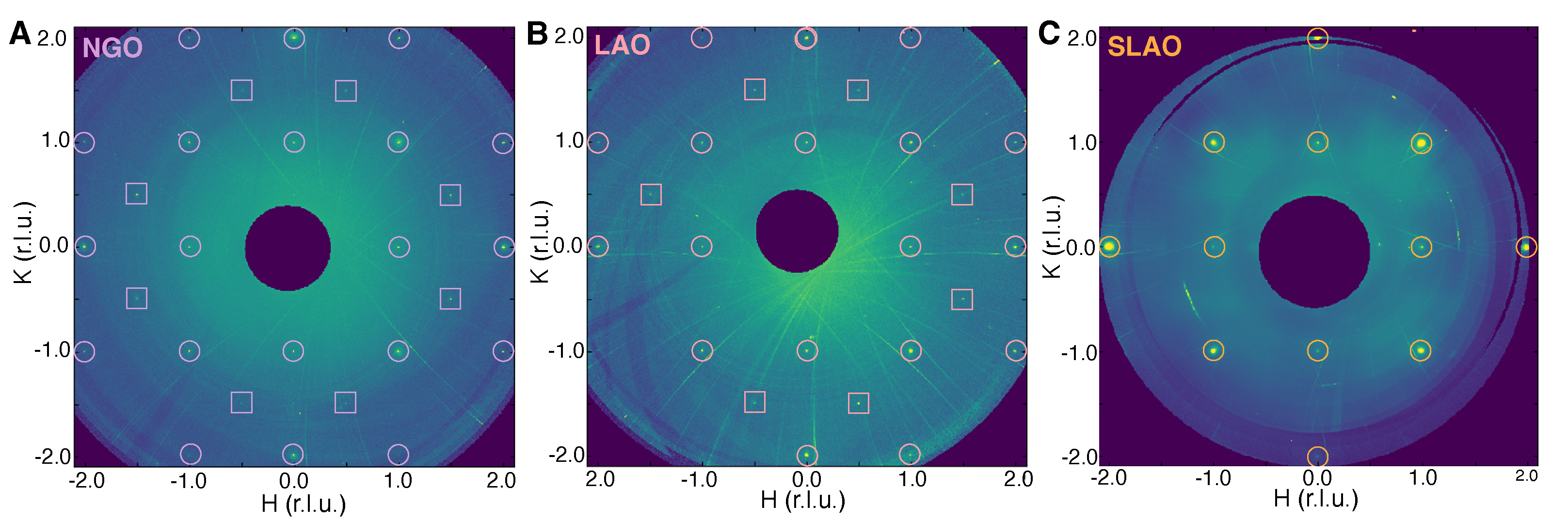}
    \caption{
    \textbf{Wider view of the $L$ = 17 HDRM cuts in Fig. 4d of the main text}. Samples presented are \Lafour films on \textbf{a)} NGO, \textbf{b)} LAO, and \textbf{c)} SLAO. Circles denote Bragg peaks due to cation ordering, and squares denote peaks arising from the in-plane octahedral rotations. Data is indexed to the \Lafour tetragonal unit cell.}
     \label{fig:supp_CHESS_1p5}
\end{figure*}

\begin{figure*}[!htb]
    \includegraphics[clip=true,width=\columnwidth]{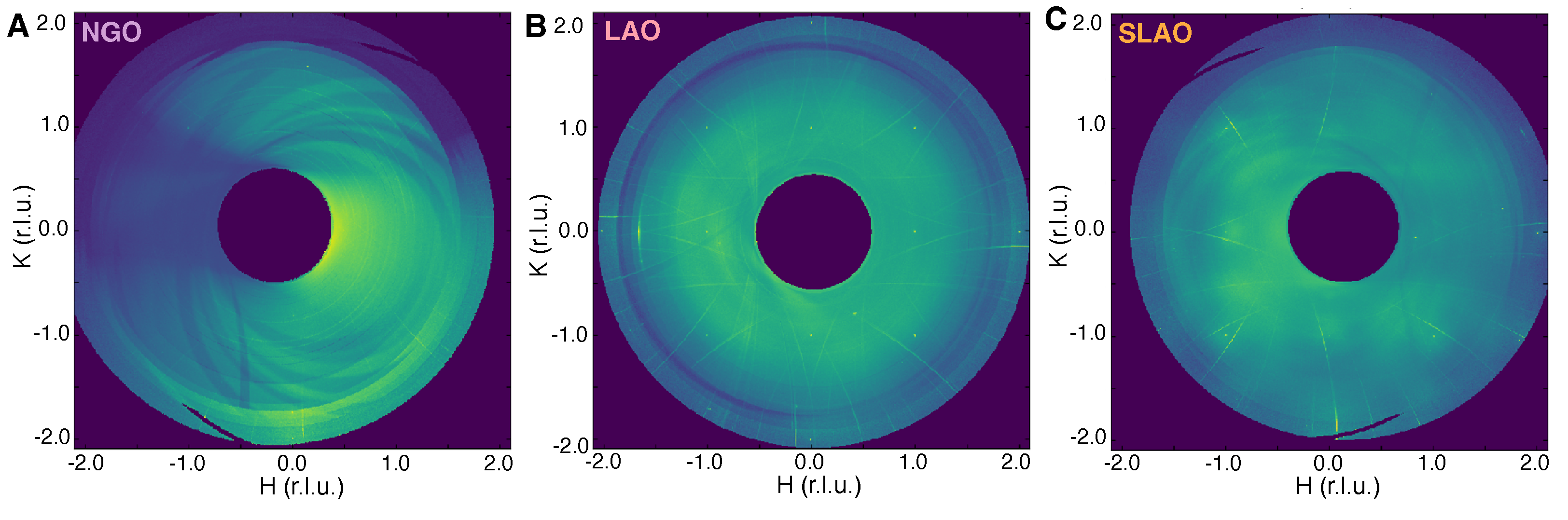}
    \caption{
    \textbf{Substrate reference HDRM, $L$ = 17.}
    \textbf{a)} NGO substrate, \textbf{b)} LAO substrate, and \textbf{c)} SLAO substrate. 
    Data is indexed to the \Lafour tetragonal unit cell.
     }\label{fig:supp_CHESS_substrates}
\end{figure*}

We include additional HDRM analysis to further corroborate the evolution of in-plane symmetry presented in Fig.~\ref{fig:COBRA}. First, we index the HDRM data to the tetragonal \Lafour unit cell to facilitate comparison with the CTR measurements, which are also indexed to the tetragonal unit cell. Wider-view cuts of those provided in Fig.~\ref{fig:COBRA}D-F are included in Fig.~\ref{fig:supp_CHESS_1p5}. These cuts at $L$ = 17 consistently show intensity at (±1.5 ±0.5 $L$) on NGO and LAO, but not on SLAO. We note that equivalent peaks are also detected (or not detected) at (±0.5 ±1.5 $L$); this $H$,$K$ equivalency is likely due to the contribution of signal from various twinned regions within the film (see Fig ~\ref{fig:supp_NGO_twin}). Equivalent HDRM scans of the substrates (Fig.~\ref{fig:supp_CHESS_substrates}) do not exhibit peaks at any half-integer $H$ or $K$, confirming that the (1.5 0.5 $L$)-type intensity in Figs.~\ref{fig:COBRA} and~\ref{fig:supp_CHESS_1p5} arises from the film itself.  
 
\begin{figure*}[!htb] \includegraphics[clip=true,width=\columnwidth]{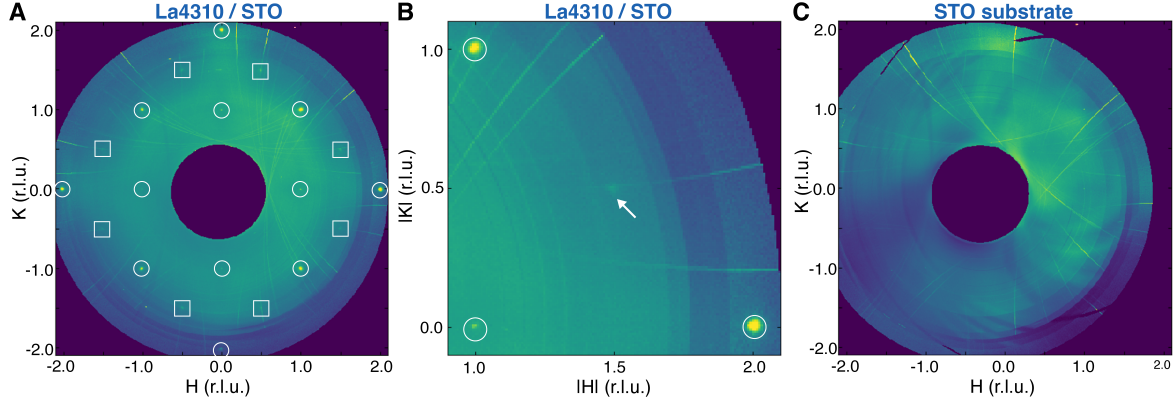}
    \caption{
    \textbf{HDRM characterization of \Lafour on STO, $L$ = 17.}
    \textbf{a)} Wide-view cut. Circles denote Bragg peaks due to cation ordering, and squares denote peaks arising from the in-plane distortions.
    \textbf{b)} Zoomed in view, with arrow pointing to an example peak arising from in-plane rotations. 
    \textbf{c)} STO substrate reference.    
    Data is indexed to the \Lafour tetragonal unit cell, and La4310 is used as an abbreviation for \Lafour. 
    }\label{fig:supp_CHESS_STO}
\end{figure*}

In addition to the three samples presented in the main text, we also perform HDRM characterization of tensile-strained \Lafour on STO (Fig.~\ref{fig:supp_CHESS_STO}). Again, we observe peaks at (±1.5 ±0.5 $L$) in addition to the peaks at integer $H$, $K$ from cation ordering, indicating that the \Lafour sample under higher tensile strain on STO ($+1.6$\%) adopts a similar in-plane symmetry to the film under mild tensile strain on NGO ($+0.3$\%).

\begin{figure*}[!htb]
    \includegraphics[clip=true,width=\columnwidth]{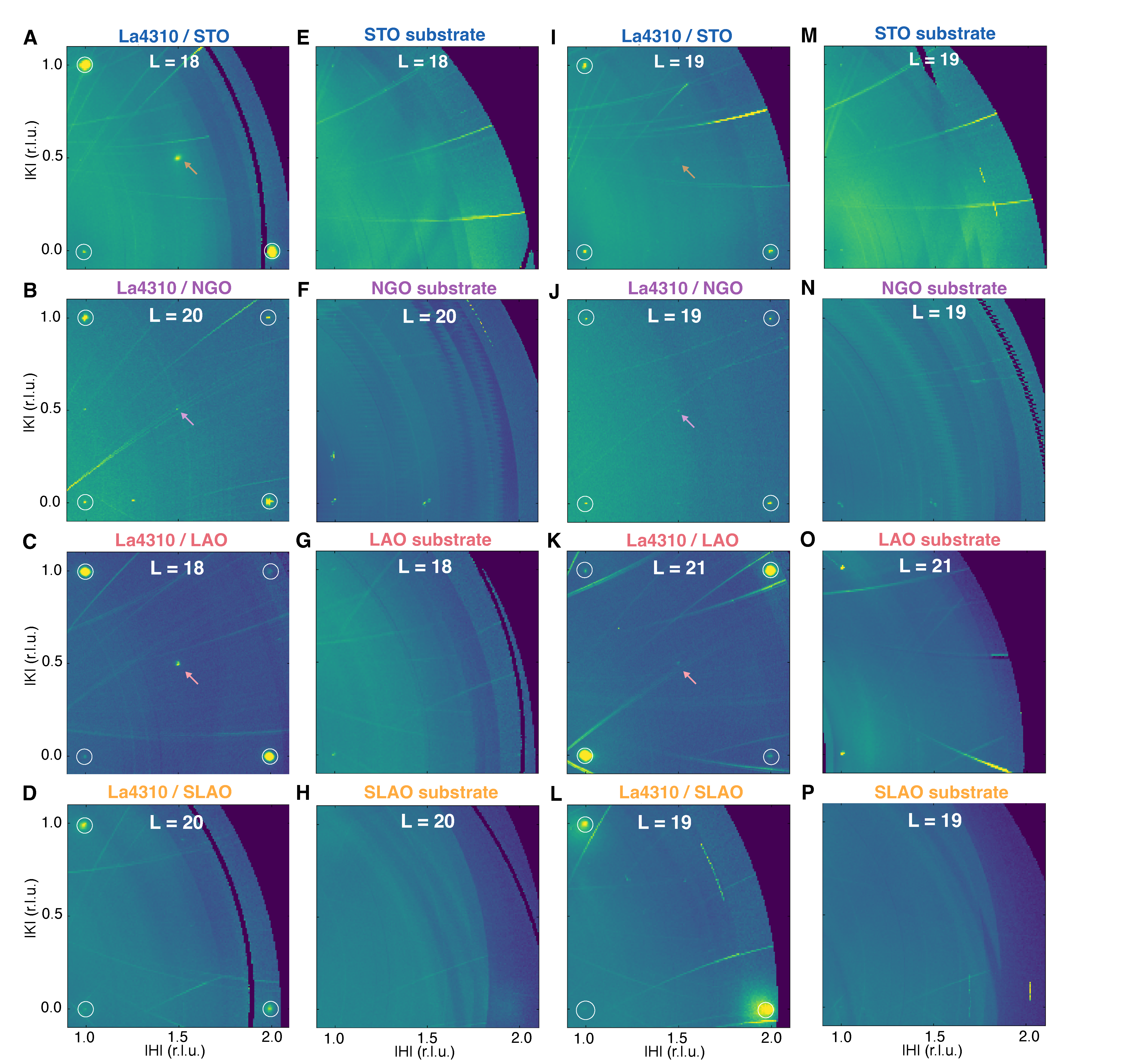}
    \caption{\textbf{HDRM cuts at additional $L$ for the \Lafour strained films}. 
    \textbf{a-d)} Cuts at even $L$ for \Lafour films on STO, NGO, LAO, and SLAO, respectively.
    \textbf{e-h)} Substrate references for the cuts in (a-d). 
    \textbf{i-l)} Cuts at odd $L$ for \Lafour films on STO, NGO, LAO, and SLAO, respectively.
    \textbf{m-p)} Substrate references for the cuts in (i-l). Circles denote Bragg peaks arising from cation ordering, and arrows indicate peaks at (1.5 0.5 $L$) arising from in-plane rotations. La4310 is used to abbreviate \Lafour. 
    }\label{fig:supp_CHESS_mega}
\end{figure*}

To verify that these trends in in-plane rotations are consistent at at several $L$, we include cuts at additional odd and even $L$ in Fig.~\ref{fig:supp_CHESS_mega}. For each substrate, the pair of odd and even $L$ presented is intentionally chosen such that there is no contribution from the substrate at the (1.5 0.5 $L$)-type peaks to isolate the contributions of the \Lafour film itself. These results further corroborate the presence of in-plane octahedral rotations on STO, NGO, and LAO -- and corresponding weakening of in-plane rotations upon the highest compressive strain on SLAO. 


\newpage
\noindent \underline{D. Low-temperature HDRM}

\begin{figure*}[!htb]
    \includegraphics[clip=true,width=\columnwidth]{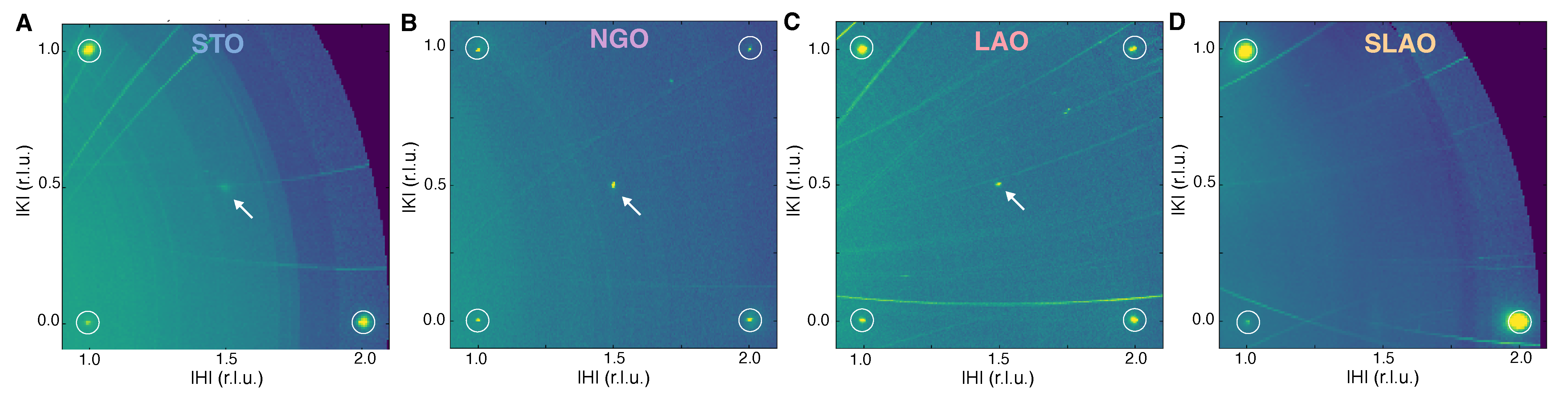}
    \caption{
    \textbf{HDRM at low temperature ($\sim60$ K), $L$ = 17.} Samples presented are \Lafour films on \textbf{a)} STO, \textbf{b)} NGO, \textbf{c)} LAO, and \textbf{d)} SLAO. Circles denote Bragg peaks due to cation ordering, and and arrows indicate peaks at (1.5 0.5 $L$) arising from in-plane rotations.
    }\label{fig:supp_CHESS_cold3p5}
\end{figure*}

We next investigate the low-temperature crystal structure to identify any potential structural changes induced by the density wave transition. The samples are cooled by flowing helium gas near the sample surface, and measurements were taken at lowest temperature possible ($\sim60$ K) while avoiding crystallization of ice on the surface of the film due to ambient humidity. The temperature is measured at the source of helium gas, so the ``real'' sample temperature is likely higher, although still well below the density wave transition temperature ($>130$ K) observed for \Lafour on STO and NGO. HDRM cuts equivalent to those presented in Fig.~\ref{fig:COBRA} and~\ref{fig:supp_CHESS_STO} for the cooled samples are shown in Fig.~\ref{fig:supp_CHESS_cold3p5}. Aside from slight variations in intensity, we detect no change in the symmetry of the in-plane rotations upon cooling below \Tdw, suggesting the space groups assigned from room-temperature data remain consistent as the sample is cooled. 

\begin{figure*}[!htb]
    \includegraphics[clip=true,width=\columnwidth]{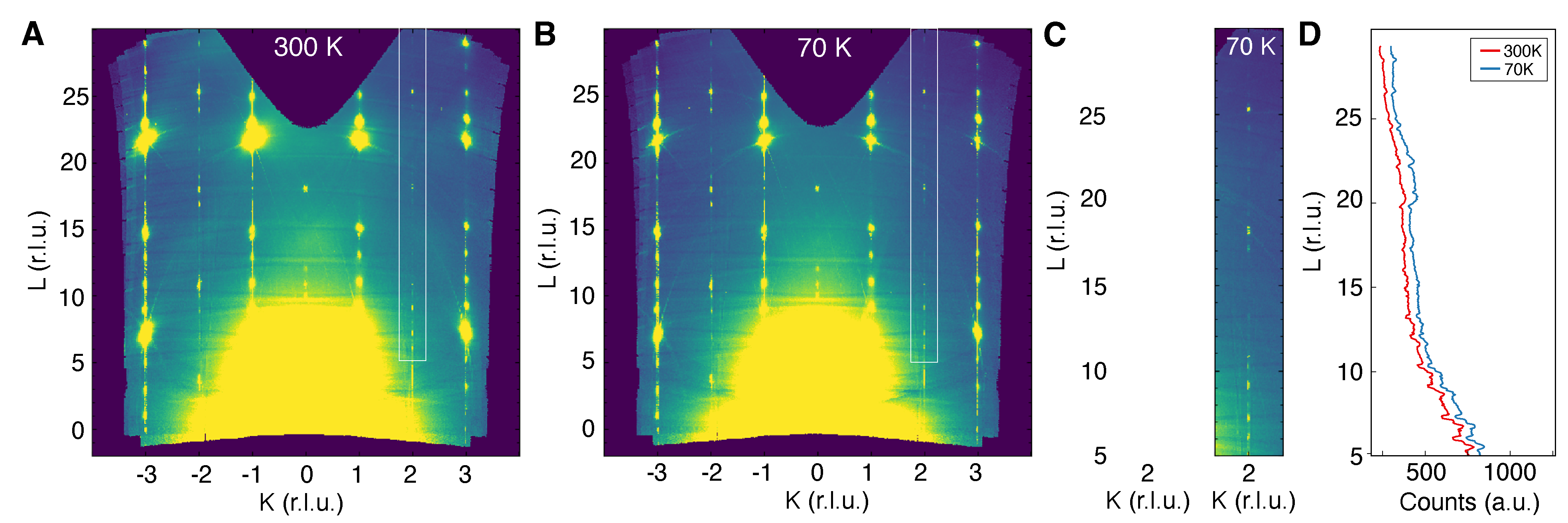}
    \caption{
    \textbf{Temperature-dependent HDRM for \Lafour on NGO, $H$ = 1.} 
    \textbf{a)} Room temperature (300 K) scan. 
    \textbf{b)} Low-temperature (70 K) scan. White boxes indicate example region where additional peaks from density wave superlattice ordering were previously detected in bulk \Lafour \cite{Zhang2020_intertwinedDW}, but we do not observe intensity in our thin films. 
    \textbf{c)} Closer view of the area around $K$ = 2 for the 300 K and 70 K datasets. 
    \textbf{d)} Line cuts around (1.75, 1, $L$), where the density wave was approximately observed in bulk crystal \Lafour \cite{Zhang2020_intertwinedDW}. The data were integrated around a window of ±0.1 in $H$ but show no additional peaks detected at low temperatures. 
    All data is indexed to the orthorhombic unit cell.
    }
    \label{fig:supp_CHESS_noCDW}
\end{figure*}

We also attempted to characterize the density wave structural distortion as previously demonstrated in bulk \cite{Zhang2020_intertwinedDW}. Because the wave vector associated with the charge component of the density wave, $q_{DW}$, is along the orthorhombic $b^*$-axis in bulk, we now index HDRM data to the orthorhombic unit cell. When comparing data for \Lafour on NGO above and below the transition temperature (Fig.~\ref{fig:supp_CHESS_noCDW}), we are unable to identify the presence of additional Bragg peaks and corresponding $q_{DW} = 0.76~b^*$ associated with the density wave transition. We choose the \Lafour film on NGO for this comparison, as this sample is the most ``bulk-like'' due to the low amount of strain imparted by the substrate (+0.3\%). Our null observation may be due to resolution and intensity limits of HDRM, as the diffracted signal is largely dominated by substrate over the small volume fraction of the thin film of interest. Given that the film Bragg peaks are already low intensity, it would be extremely challenging to detect even weaker peaks arising from density wave superlattice ordering, as these superlattice peaks are reported to be $10^4$ times lower intensity than the strongest Bragg peaks \cite{Zhang2020_intertwinedDW}. 
Improvements in X-ray scattering experimentation, either via higher resolution or longer count times, to resolve weak off-Bragg peak intensity in thin films would be an exciting future direction. 

\renewcommand{\citenumfont}[1]{S#1}

\clearpage
\newpage
\subsection{Detailed visualization of \Lafour crystal structures}
\label{sec:supp_structures}


\begin{figure}[!htb]
    \includegraphics[width=0.7\linewidth]{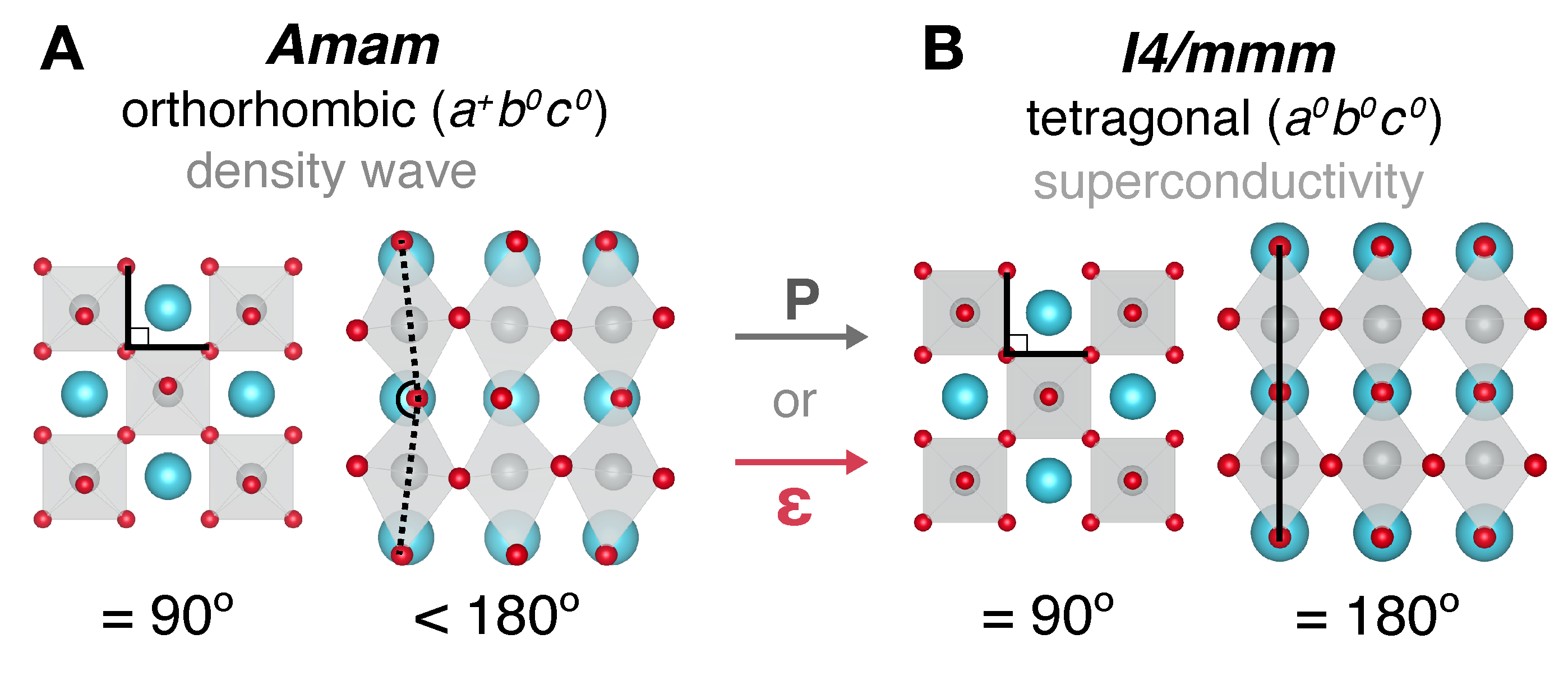}
    \caption{\textbf{Structural evolution of \LNO for comparison.} In this bilayer RP nickelate system, both hydrostatic pressure and epitaxial strain induce a direct transition to the highest symmetry $I4/mmm$ structure.
    \textbf{a)} $Amam$ \LNO at ambient pressure adapted from \cite{Ling2000La3Ni2O7}. 
    \textbf{b)} \LNO in the tetragonal $I4/mmm$ space group \cite{Sun2023_La327} found to be superconducting both in bulk under hydrostatic pressure and in thin films under compressive epitaxial strain.}
    \label{fig:ext_La327_structure}
\end{figure}

\begin{figure*}[!htb]
    \includegraphics[clip=true,width=0.8\columnwidth]{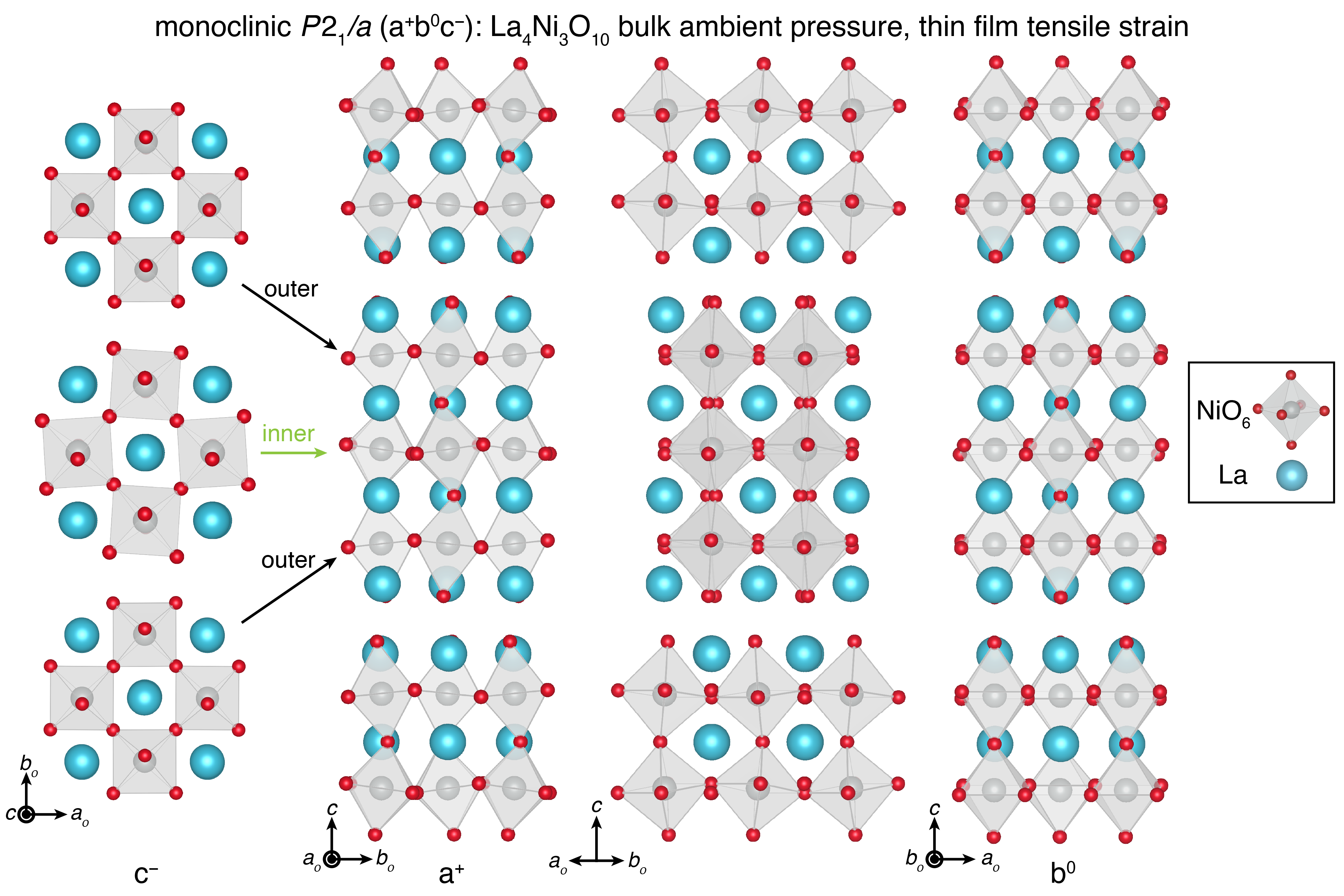}
    \caption{
    \textbf{Monoclinic $P2_1/a$ structure} from bulk ambient pressure refinement \cite{Zhang2020_LaPr4310floatingzone}, and also stabilized in \Lafour on tensile strain in our work. Pseudo-orthorhombic (o) indexing is used.}
    \label{fig:supp_p21a}
\end{figure*}

\begin{figure*}[!htb]
    \includegraphics[clip=true,width=0.8\columnwidth]{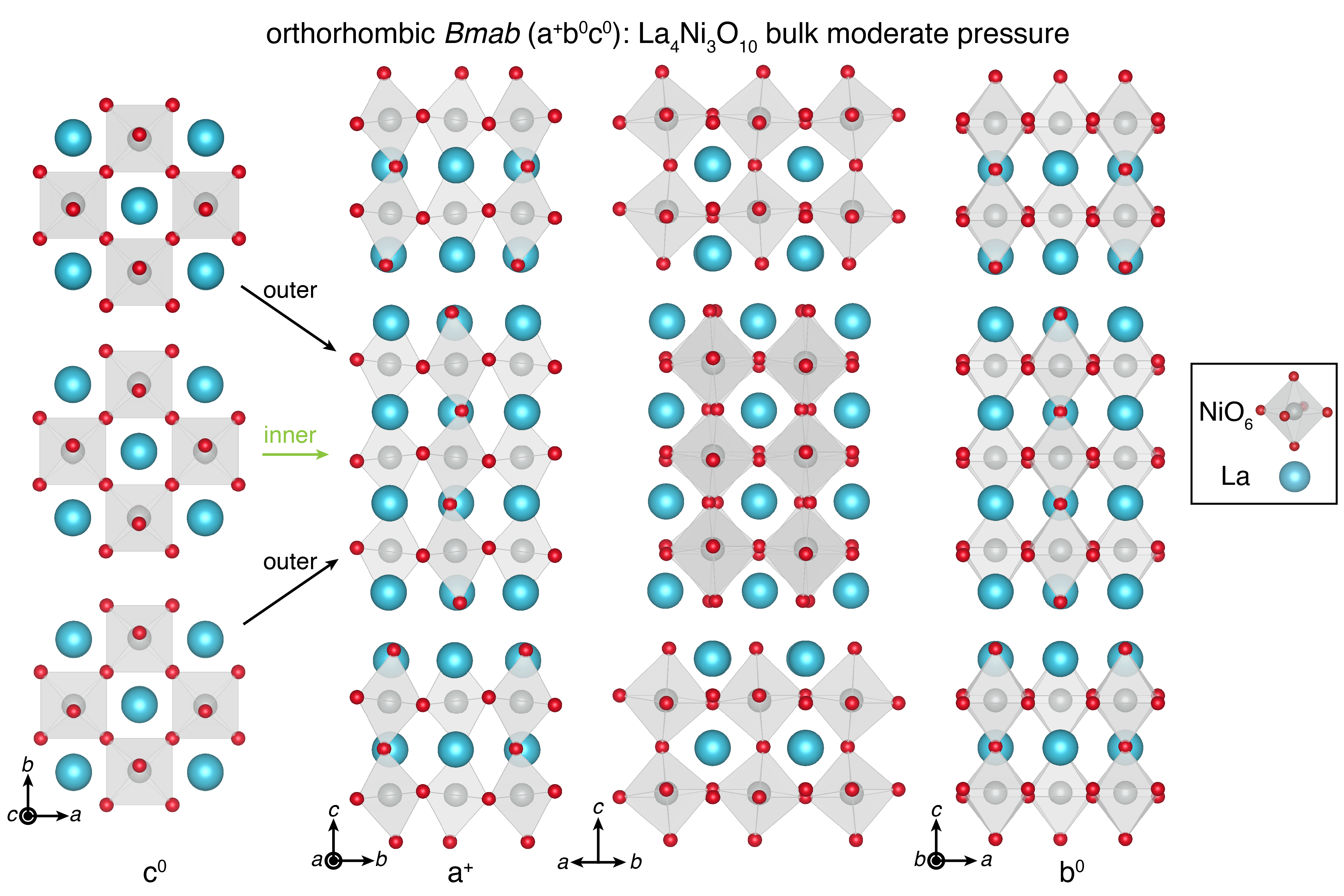}
    \caption{
    \textbf{Orthorhombic $Bmab$ structure} observed in bulk under moderate pressure \cite{Li2025_La4310structure}, with structure file from bulk refinement \cite{Ling2000_RP123neutron}.}
    \label{fig:supp_bmab}
\end{figure*}

\begin{figure*}[!htb]
    \includegraphics[clip=true,width=0.8\columnwidth]{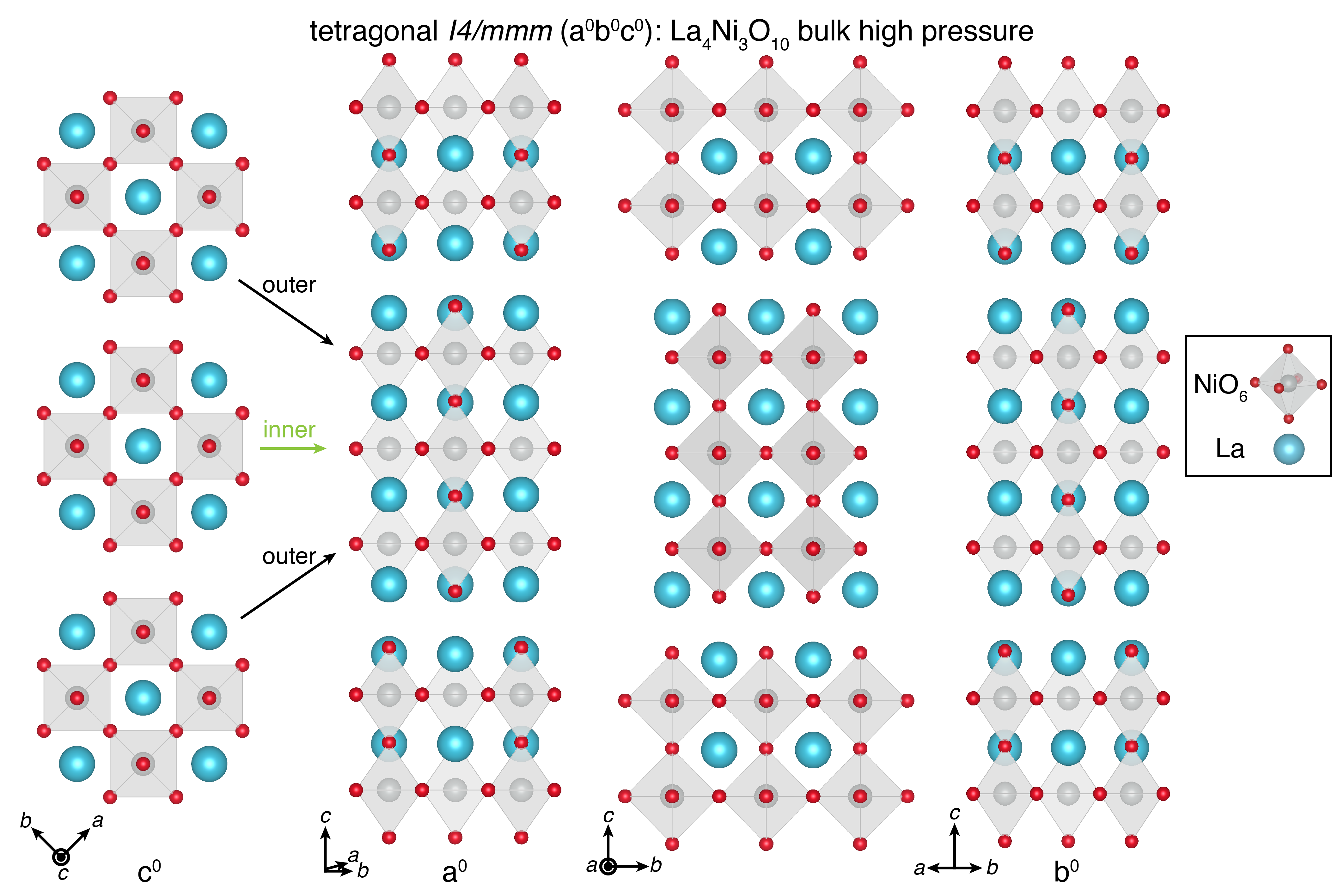}
    \caption{
    \textbf{Tetragonal $I4/mmm$ structure} observed in bulk under large pressures in the superconducting state, with structure file from bulk refinement \cite{Nagell2017_LaCo4310}.}
    \label{fig:supp_i4mmm}
\end{figure*}

\begin{figure*}[!htb]
    \includegraphics[clip=true,width=0.8\columnwidth]{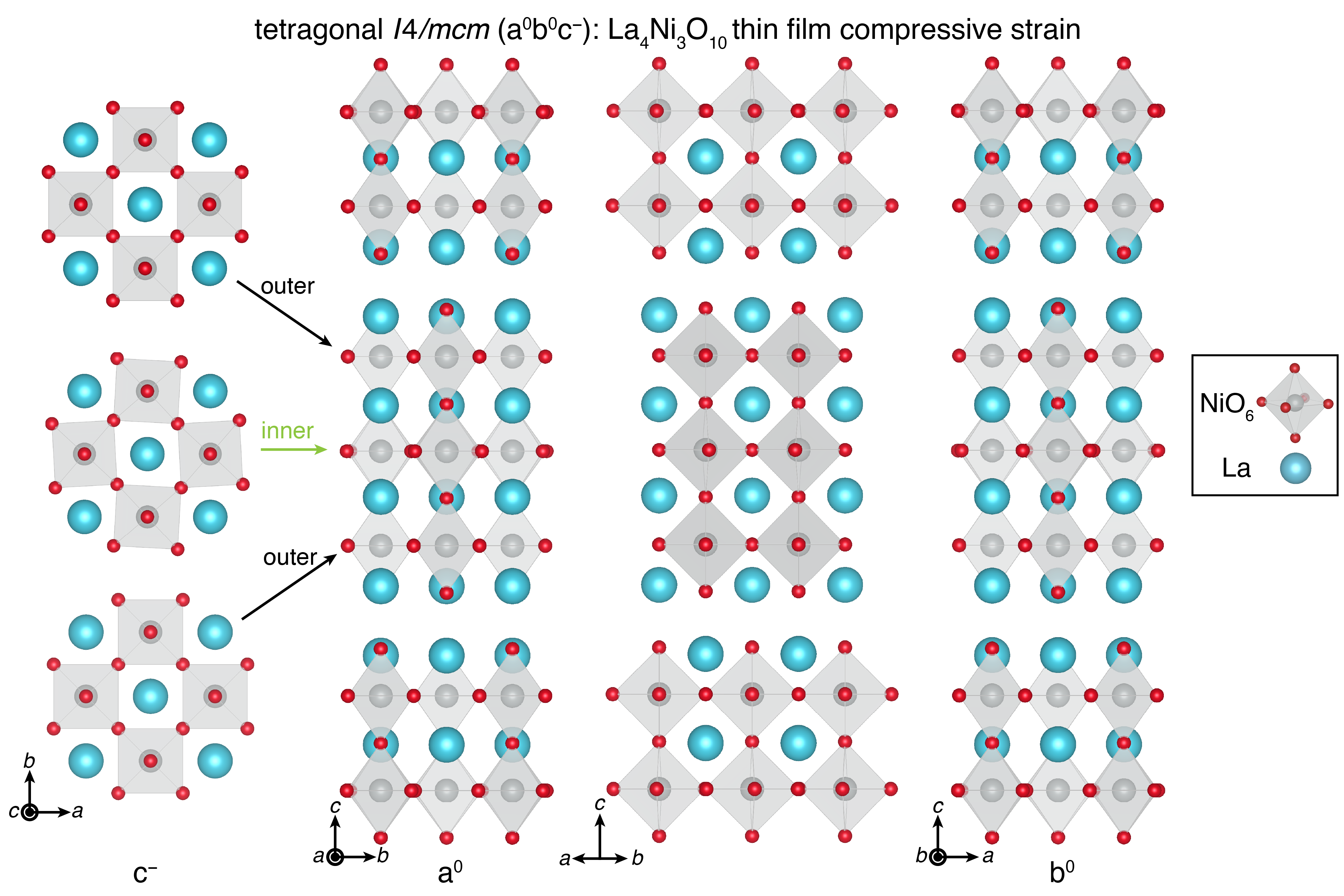}
    \caption{
    \textbf{Tetragonal $I4/mcm$ structure} observed in \Lafour on compressive strain in our work.}
    \label{fig:supp_i4mcm}
\end{figure*}

\clearpage
\newpage
\subsection{Substrate preparation}
\label{sec:supp_SLAOsubs}

\begin{figure*}[!htb]
    \includegraphics[clip=true,width=1.0\columnwidth]{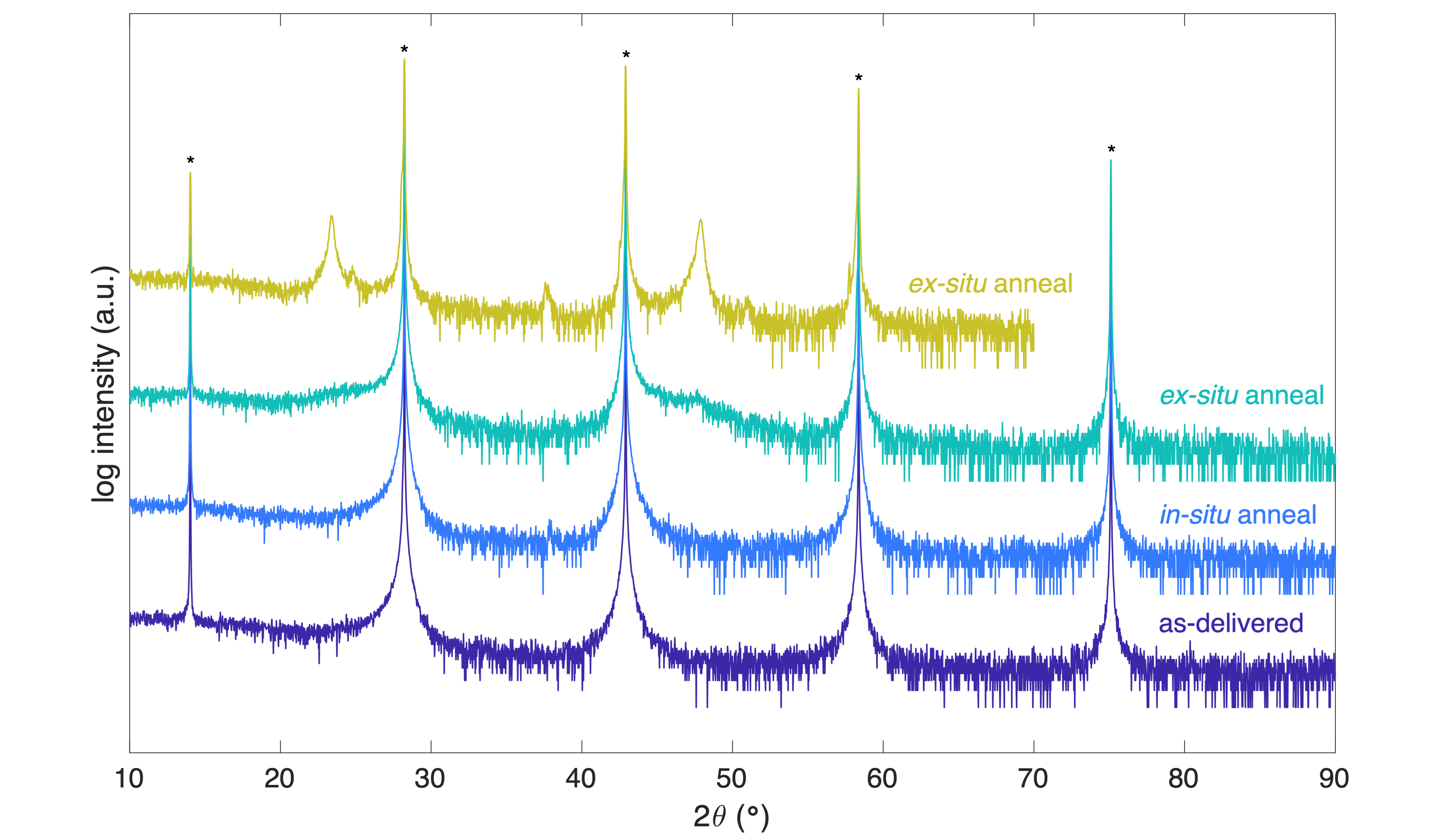}
    \caption{
    \textbf{Preparation of SLAO substrates.} XRD of as-delivered and \textit{in-situ} MBE-annealed SLAO shows minimal degradation and secondary phase formation, while SLAO annealed \textit{ex-situ} in air using previously reported preparation methods \cite{Kim2022_A2BO4} demonstrates significant degradation and secondary phases. Asterisks denote substrate peaks of SLAO(001).
}
    \label{fig:supp_SLAO_XRD}
\end{figure*}
\begin{figure*}[!htb]
    \includegraphics[clip=true,width=1.0\columnwidth]{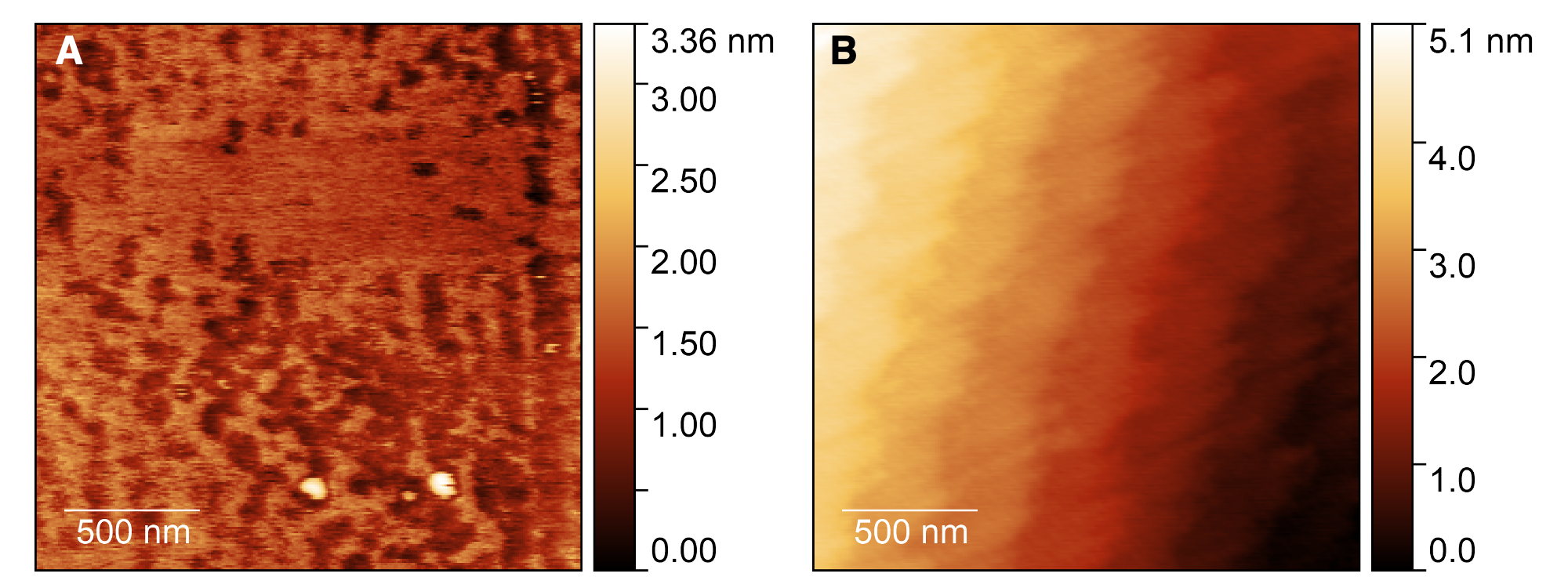}
    \caption{
    \textbf{AFM of SLAO substrate.} \textbf{a)} As-delivered and \textbf{b)} \textit{in-situ} MBE annealed substrate up to $\sim$550 \textdegree C in $\sim$1E-7 torr distilled ozone, showing significant improvement in surface morphology after \textit{in-situ} anneal.}
    \label{fig:supp_SLAO_AFM}
\end{figure*}

\clearpage
\newpage
\putbib[supplement]
\end{bibunit}

\end{document}